\documentclass[fleqn,usenatbib]{mnras}

\usepackage{newtxtext,newtxmath}

\usepackage[T1]{fontenc}

\DeclareRobustCommand{\VAN}[3]{#2}
\let\VANthebibliography\thebibliography
\def\thebibliography{\DeclareRobustCommand{\VAN}[3]{##3}\VANthebibliography}

\usepackage{graphicx}	
\usepackage{amsmath}	

\newcommand{\be}{\begin{equation}}
\newcommand{\ee}{\end{equation}}
\newcommand{\bx}{\mathbf{x}}
\newcommand{\bk}{\mathbf{k}}
\newcommand{\p}{\mathcal{P}}

\newcommand{\ftp}{\widetilde{\p}}
\newcommand{\ft}{\widetilde{T}}
\newcommand{\fb}{\widetilde{B}}
\newcommand{\fdel}{\widetilde{\delta}}
\newcommand{\rij}{\mathcal{R}_{ij}}
\newcommand{\di}[2]{d_{#1}^{(#2)}}

\title[Characteristic Functions for Cosmology]{Characteristic Functions for Cosmological Cross-Correlations}

\author[P.~C.~Breysse et al.]{
Patrick C.~Breysse,$^{1}$\thanks{E-mail: pb2555@nyu.edu (PCB)}
Dongwoo T.~Chung,$^{2,3}$
H\aa vard T.~Ihle$^{4}$
\\
$^{1}$Center for Cosmology and Particle Physics, Department of Physics, New York University, 726 Broadway, New York, NY, 10003, USA\\
$^{2}$Canadian Institute for Theoretical Astrophysics, University of Toronto, 60 St. George Street, Toronto, ON M5S 3H8, Canada\\
$^{3}$Dunlap Institute for Astronomy and Astrophysics, University of Toronto, 50 St. George Street, Toronto, ON M5S 3H4, Canada\\
$^{4}$Institute of Theoretical Astrophysics, University of Oslo, P.O. Box 1029 Blindern, N-0315 Oslo, Norway\\
}

\date{Accepted XXX. Received YYY; in original form ZZZ}

\pubyear{2022}

\begin{document}
\label{firstpage}
\pagerange{\pageref{firstpage}--\pageref{lastpage}}
\maketitle

\begin{abstract}
We introduce a novel unbiased, cross-correlation estimator for the one-point statistics of cosmological random fields.  One-point statistics are a useful tool for analysis of highly non-Gaussian density fields, while cross-correlations provide a powerful method for combining information from pairs of fields and separating them from noise and systematics.  We derive a new Deconvolved Distribution Estimator that combines the useful properties of these two methods into one statistic.  Using two example models of a toy Gaussian random field and a line intensity mapping survey, we demonstrate these properties quantitatively and show that the DDE can be used for inference.  This new estimator can be applied to any pair of overlapping, non-Gaussian cosmological observations, including large-scale structure, the Sunyaev-Zeldovich effect, weak lensing, and many others.
\end{abstract}




\section{Introduction}

At the most fundamental level, the goal of observational cosmology is to measure the statistics of large-scale density fields.  This general concept takes many forms, from intensity fluctuations of the cosmic microwave background (CMB) \citep{Smoot1992,Hinshaw2013,Planck2020a} to galaxy counts tracing their underlying dark matter \citep{Eisenstein2005,Alam2021} to many other similar observables.  These measurements have led to the $\Lambda$-Cold Dark Matter ($\Lambda$CDM) concordance model of cosmology, in which the statistics of these fields are described by a simple six-parameter model \citep{Hamana2020,Planck2020,Dutcher2021,Abbott2022,Aiola2022,Dvornic2022}.

The principal tool for most of these analyses is the power spectrum, which quantifies how the Fourier transform of a density field evolves with spacial scale.  For fields like the CMB, which has been shown to be nearly perfectly isotropic, homogeneous, and Gaussian \citep{Planck2020b}, the power spectrum fully describes the statistics of the field.  As with a one-dimensional Gaussian distribution, all higher-order moments of a Gaussian random field are either zero or expressible in terms of the two-point power spectrum.  At early times, the large-scale matter density structure of the universe is fairly Gaussian as well, but at later times nonlinear gravitational evolution drives the production of significant small-scale non-Gaussianity \citep{Bernardeau2002}.  Baryonic effects can drive even more nonlinearity in some observables \citep{Breysse2017}.

Many tools exist to understand cosmological non-Gaussianity. Often non-Gaussianities are sought using higher-order $n$-point functions such as bispectra and trispectra \citep[e.g.][]{Bartolo2004,PlanckIX,Dizgah2021}.  For non-Gaussian fields, these quantities pick up independently non-zero values.  In this paper, though, we discuss a simpler quantity: a field's \emph{one-point statistics}.  Specifically, the probability distribution function (PDF) of a field taking a given value at a single point.  The one-point PDF summarizes information from all of the $n$-point moments in one convenient distribution \citep{Barcons1992}.  One-point distributions have been shown to be useful probes of cosmological physics, including primordial non-Gaussianity \citep{Valageas2002,Uhlemann2018,Friedrich2020}, gravitational lensing \citep{Liu2016,Patton2017,Liu2019,Thiele2020}, modified gravity \citep{Li2012,Brax2012,Cataneo2022}, the Sunyaev-Zeldovich effect \citep{Hill2014,Thiele2019}, decaying or annihilating dark matter \citep{Lee2015,Bernal2021a}, the reionization-era intergalactic medium \citep{Barkana2008,Shimabukuro2015}, the cosmic neutrino background \citep{Feyereisen2017,Bernal2021}, and unresolved galactic emission \citep{Glenn2010,Breysse2016,Breysse2017,Ihle2019,Leicht2019,Breysse2022bias}.

For all their utility, one-point methods to date lack a key feature of two- and higher point statistics: the ability to cross-correlate multiple observables.  Cross-correlations are a hallmark of modern cosmology, which project out correlated information between two overlapping measurements.  Cross-correlations lead to confident detections of faint signals \cite[e.g.][]{Vielva2006,Chang2010,Hanson2013,Tanimura2021}, as one can often arrange a situation where two observations have signals which are correlated and noise and systematics which are not.  They can also add unique information about the relationship between two fields \citep{Seljak2009,Lidz2011}.  In this paper, we therefore have a simple goal: to derive an estimator which carries the benefits of both the one-point PDF and the cross-correlation.  We show that the result, which we term the Deconvolved Distribution Estimator (DDE), carries both of these properties.

A first attempt at such an estimator is the Conditional Voxel Intensity Distribution (CVID) described in \citet{Breysse2019}.  The CVID estimator was designed for the specific case of cross-correlating 21 cm intensity maps with spectroscopic galaxy surveys \citep{Chang2010,Switzer2013,Anderson2018,Cunnington2022,Wolz2022}.  With the DDE, we generalize the \citet{Breysse2019} result into a form which can be applied to any pair of non-Gaussian random fields.  The primary mechanism by which the DDE acts is through the characteristic function, or the Fourier transform of the one-point PDF.  We show that, by manipulating the characteristic functions of a pair of fields, we can construct a quantity which only depends on their correlated components.

Since this application of characteristic functions appears relatively novel, we go on to demonstrate using a pair of example models that the DDE does carry the utility of a cross-correlation.  Our first model is is deliberately as simple as possible, the correlation of a pair of Gaussian random fields.  Though this situation would be fully described by its power spectra, it enables us to study the DDE and its measurement in a fully analytic form.  As an actual potential use-case, we imagine applying the DDE to a line intensity mapping (LIM) survey \citep{Kovetz2017,Bernal2022,esVII} consisting of the smoothed-over aggregate emission of a population of galaxies.  This is a realistic near-term application of one-point statistics, and we show that the DDE performs its job in this case as well.  Our LIM model here is still fairly simplified for pedagogical purposes, a concurrent paper by \citet{Chung2022} studies a LIM DDE problem in greater detail.

The organization of this paper is as follows: We introduce the DDE formalism in Section \ref{sec:formalism}, we build and test our toy Gaussian model in Section \ref{sec:toy}, and our LIM model in Section \ref{sec:lim}.  We discuss the results of these examples in Section \ref{sec:discussion}, and conclude in Section \ref{sec:conclusion}.  Useful mathematical details can be found in the Appendix.  Where relevant, we assume a flat $\Lambda$CDM cosmology consistent with the Planck 2018 results and assume that any distances carry an implicit factor of $h^{-1}$.  We adopt the conventions
\be
F(\bk)=\int f(\bx)e^{-i\bk\cdot\bx}d^n\bx,\ \ \ \ \ \ f(\bx)=\frac{1}{(2\pi)^n}\int F(\bk)e^{i\bk\cdot\bx}d^n{\bk},
\ee
for the Fourier transform and its inverse.

\section{Formalism}
\label{sec:formalism}

Consider a common cosmological scenario in which we have observed two correlated density fields $\delta_1(\mathbf{x})$ and $\delta_2(\mathbf{x})$ over an overlapping volume $V_{\rm surv}$.  Typically, our first step upon acquiring these data would be to compute the power spectra
\be
P_1(k)=\frac{1}{V_{\rm surv}}\left<\left|\delta_1(\bk)\right|^2\right>,\ \ \ \ P_2(k)=\frac{1}{V_{\rm surv}}\left<\left|\delta_2(\bk)\right|^2\right>,
\label{autospec}
\ee
and the cross-spectrum
\be
P_\times(k)=\frac{1}{V_{\rm surv}}\left<\delta_1(\bk)\delta_2^*(-\bk)\right>,
\label{crossspec}
\ee
where $\delta(\bk)$ indicates the spatial Fourier transform of $\delta(\bx)$.  If $\delta_1$ and $\delta_2$ are perfectly Gaussian random fields, then they are entirely described by Equations (\ref{autospec}) and (\ref{crossspec}).

However, if the target fields are non-Gaussian, there will be information contained within them that is not accessible to the power spectrum alone.  Here we seek to access this non-Gaussianity using the one-point statistics of a map (as opposed to the two-point or higher statistics probed by the power spectrum, bispectrum, etc.).  Specifically, we examine the probability distribution\footnote{To avoid confusion, we use $P$ to refer to power spectra and $\p$ to refer to probability distributions throughout} $\p(\delta)$ of a field taking a value between $\delta$ and $\delta+d\delta$ at a given point.  In practice, $\p(\delta)$ is usually estimated by gridding a map of $\delta$ into discrete pixels or voxels then computing the histogram
\be
B_i=N_{\rm vox}\int_{\delta_i}^{\delta_i+\Delta\delta}\p(\delta)d\delta,
\label{histint}
\ee
where $N_{\rm vox}$ is the total number of volume elements (here assumed to be 3D voxels) and $\Delta\delta$ is the width of a histogram bin with its lower edge at $\delta_i$.  In general, Equation (\ref{histint}) defines the expectation value of an observed histogram, where an estimate from data will have some uncertainty.  Since we are not dealing with real data in this paper, our modeled histograms and their expectation value will be identical, and we will use $B_i$ for both.

The one-point PDF $\p(\delta)$, as its name suggests, only uses information from a single density field.  Our goal here is to create an analogue of the cross-power spectrum estimator which has the following properties:
\begin{itemize}
\item Retains the non-Gaussian information from $\p(\delta)$.
\item Probes the relationship between $\delta_1$ and $\delta_2$, yielding additional information beyond the individual PDFs $\p(\delta_1)$ and $\p(\delta_2)$.
\item Is unbiased by any noise or systematic effects which are uncorrelated between the $\delta_1$ and $\delta_2$ measurements.
\end{itemize}
One obvious way to satisfy the first two requirements would be to examine the joint two-dimensional probability distribution $\p(\delta_1,\delta_2)$ between the two fields, or similarly the two-dimensional histogram
\be
B_{ij}=N_{\rm vox}\int_{\delta_{1,i}}^{\delta_{1,i}+\Delta\delta_1}\int_{\delta_{2,j}}^{\delta_{2,j}+\Delta\delta_2}\p(\delta_1,\delta_2)d\delta_1d\delta_2.
\ee
This quantity is clearly sensitive to the non-Gaussian correlation between the two fields.  Crucially, though, it lacks the unbiased quality of Eq. (\ref{crossspec}).  If we want a true cross-spectrum equivalent, we need to work with the \emph{characteristic functions} of our density fields.

To demonstrate why, we rely on two facts:  First, that the PDF of the sum of two independent variables is the convolution of the two PDFs.  In other words, if we observe some signal $\delta^{\rm obs}$ which is the sum of a target signal $\delta^S$ and an independent noise $\delta^N$, our observed PDF will be
\be
\p^{\rm obs}(\delta)=\p^S\circ\p^N(\delta),
\label{pconv}
\ee
and the observed joint 2D PDF will be
\be
\p_{\rm 2D}^{\rm obs}(\delta_1,\delta_2)=\p_{\rm 2D}^S\circ\p_{\rm 2D}^N(\delta_1,\delta_2).
\label{pconv2d}
\ee
The $\circ$ operator denotes the operation
\be
\p^S\circ\p^N(\delta)=\int\p^S(\delta')\p^N(\delta-\delta')d\delta'.
\ee
Our second key fact is the Fourier convolution theorem, which states that the Fourier transform of a convolution of two distributions is the product of their Fourier transforms.  Thus, if we Fourier transform both sides of Equations (\ref{pconv}) and (\ref{pconv2d}), we get
\be
\ftp^{\rm obs}(\fdel)=\ftp^S(\fdel)\ftp^N(\fdel),
\label{ft1d}
\ee
and
\be
\begin{aligned}
\ftp_{\rm 2D}^{\rm obs}(\fdel_1,\fdel_2)&=\ftp_{\rm 2D}^S(\fdel_1,\fdel_2)\ftp_{\rm 2D}^N(\fdel_1,\fdel_2)\\&=\ftp_{\rm 2D}^S(\fdel_1,\fdel_2)\ftp^N_1(\fdel_1)\ftp^N_2(\fdel_2),
\end{aligned}
\label{ft2d}
\ee
where $\ftp(\fdel)$ is the characteristic function of the field $\delta(\bx)$, obtained by Fourier transforming the probability distribution $\p(\delta)$, and $\fdel$ is the Fourier conjugate of $\delta$.  Note that we are working with the Fourier transform of the probability distribution itself, not the spatial Fourier transform of $\delta(\bx)$ used to compute the power spectrum.  The second equality in Eq. (\ref{ft2d}) is obtained by assuming, as is typical in cross-correlations, that any noise or other systematic contribution to $\delta^{\rm obs}$ is uncorrelated between the two fields.

With Equations (\ref{ft1d}) and (\ref{ft2d}) in hand, we can now construct our cross-correlation estimator.  We define the quantity
\be
\mathcal{R}(\fdel_1,\fdel_2)\equiv \frac{\ftp_{\rm 2D}^{\rm obs}(\fdel_1,\fdel_2)}{\ftp^{\rm obs}_1(\fdel_1)\ftp^{\rm obs}_2(\fdel_2)}-1.
\label{Rdef}
\ee
To see why we have chosen this specific ratio, let us insert Equations (\ref{ft1d}) and (\ref{ft2d}) into our definition of $\mathcal{R}$:
\be
\begin{aligned}
\mathcal{R}(\fdel_1,\fdel_2)&=\frac{\ftp_{\rm 2D}^S(\fdel_1,\fdel_2)\ftp^N_1(\fdel_1)\ftp^N_2(\fdel_2)}{\ftp^S_1(\fdel_1)\ftp^N_1(\fdel_1)\ftp^S_2(\fdel_2)\ftp^N_2(\fdel_2)}-1\\
&=\frac{\ftp_{\rm 2D}^{S}(\fdel_1,\fdel_2)}{\ftp^{S}_1(\fdel_1)\ftp^{S}_2(\fdel_2)}-1.
\end{aligned}
\label{ddeproof}
\ee
By moving to characteristic function space we have turned our noise convolutions into products which appear identically in the numerator and denominator, and thus the noise contribution cancels out and we are left with a quantity which depends only on signal.  We subtract one from $\mathcal{R}$ so that $\mathcal{R}=0$ in the case where one or both maps are entirely noise dominated

In practice, when we estimate the PDFs using histograms, we will construct the estimator 
\be
\rij=N_{\rm{vox}}\frac{\fb_{ij}}{\fb_{1,i}\fb_{2,j}}-1,
\label{dde}
\ee
where $\fb_i$ is the Fourier transform of the histogram $B_i$.  By the same arguments as above, we can see that
\be
\left<\rij\right>\approx \mathcal{R}(\fdel_{1,i},\fdel_{2,j}),
\ee
making $\rij$ a (mostly) unbiased estimator of $\mathcal{R}$.  The estimator is only mostly unbiased because the finite bin size necessary to define a histogram renders the equality inexact.  We have found this binning error to be small for all computations used for this work, but care should be taken as this has not been proven for the general case and it may be necessary to apply a correction in some cases \citep[e.g.][]{Sun2022}.  The factor of $N_{\rm vox}$ in Eq. (\ref{dde}) is a necessary normalization to set the amplitude of the histogram estimator to that computed from the unbinned characteristic functions.

We will refer to the $\rij$ estimator as the Deconvolved Distribution Estimator (DDE).  From Eq. (\ref{ddeproof}), it is clear that the DDE meets the third of our cross-correlation criteria above.  So long as any noise or systematics present in each observation is uncorrelated both with the individual signals and between the two data sets, $\rij$ will be an unbiased quantity dependent only on the two signals.  It is also clear, through the presence of $\p_S(\delta_1,\delta_2)$, that the DDE carries information about the relationship between the two fields beyond what is present in either single-field PDF, satisfying our second criterion.  We dedicate most of the rest of this work to demonstrating that the DDE fulfills the first criterion, that it actually contains usable information.

Answering this question in detail will require quantitative demonstrations in the next pair of sections, but we can glean some insights into the behavior of the DDE through inspection.  Because we construct it from characteristic functions, $\rij$ will in general be a complex number, so we will need to keep track of both its real and imaginary parts.  Because the original histograms are by definition real, however, only half of the possible $\fdel$ bins will be independent.  We will generally account for this by only considering bins with $\fdel_2>0$.  In fact, we will neglect all bins in which $\fdel_1=0$ or $\fdel_2=0$, as, assuming the distributions are properly normalized, all of these bins will be identically zero.  Finally, the DDE will be unaffected by any global, additive constant applied to either field.  Such a constant will enter the characteristic function as a phase, which will appear equally in the numerator and denominator and thus cancel out.  The DDE is thus insensitive to the global mean of either field, though of course many cosmological measurements are already insensitive to the global mean regardless.

As with the standard cross-spectrum, the DDE can be applied to any pair of correlated fields with uncorrelated noise.  For our demonstrations here, we will make the simplifying assumption that the two signal fields are identical, with independent noise.  This is an approximation of the general cross-correlation case, but it is also an exact model of a situation which arises commonly in cosmological data analysis.  When measuring an autocorrelation, any unaccounted-for systematics in a data set will bias the measurement.  A way around this problem is to split a data set into two halves with each half mapping the same signal with independent noise.  For example, a hypothetical two-year survey could separate maps made from its first year of data from maps made from its second.  Then the two subsets can be cross-correlated.  The expectation value of this cross-correlation will be the same as the auto-correlation, but uncorrelated systematics will enter as excess error rather than bias.

In the case of identical signals, we can simplify the expected form of the DDE.  The one-dimensional distributions of our two signals are identical, with $\p_1(\delta)=\p_2(\delta)$.  The two-dimensional distribution takes the diagonal form
\be
\p_{\rm 2D}(\delta_1,\delta_2)=\p_1(\delta)\delta_D(\delta_1-\delta_2),
\ee
where $\delta_D$ is the Dirac delta function. As with the power spectrum, many types of systematics can bias $\p_1^{\rm obs}$, but if we can split the data into portions with independent noise our estimate of $\mathcal{R}$ will be unbiased.  Under these assumptions, that estimation will have the expectation value
\be
\mathcal{R}(\fdel_1,\fdel_2)=\frac{\ftp_1(\fdel_1+\fdel_2)}{\ftp_1(\fdel_1)\ftp_1(\fdel_2)}-1.
\label{Rsignal}
\ee
Our specific aim for the remainder of this work will be to quantify how much of the information content of an unbiased $\p_1(\delta)$ measurement is retained in this DDE.  An exploration of the more general case of cross-correlating two different fields can be found in \citet{Chung2022}.

\subsection{Error analysis}
Though the noise does not bias $\rij$, it does still contribute to the error.  If we treat each map voxel as an independent draw from an underlying distribution, the errors on each histogram bin will have a multinomial distribution \citep{Ihle2019}.  A one-dimensional histogram will have covariance
\be
C_{ij}^{\rm 1D}=\left<B_iB_j\right>-\left<B_i\right>\left<B_j\right>=B_i\delta_{ij}^K-\frac{1}{N_{\rm vox}}B_iB_j.
\label{cov1}
\ee
The off-diagonal covariance comes from the fact that each histogram must sum to exactly $N_{\rm vox}$, and $\delta_{ij}^K$ is the Kronecker delta function. A similar result holds for the 2-D histogram
\be
C^{\rm 2D}_{ijk\ell}=\left<B_{ij}B_{k\ell}\right>-\left<B_{ij}\right>\left<B_{k\ell}\right>=B_{ij}\delta_{ik}^K\delta_{j\ell}^K-\frac{1}{N_{\rm vox}}B_{ij}B_{k\ell}.
\label{cov2}
\ee
We also need to account for the fact that the the 1-D and 2-D histograms will be correlated with each other as
\be
C^{\rm 1D\times2D}_{ijk}=\left<B_iB_{jk}\right>-\left<B_i\right>\left<B_{jk}\right>=B_{jk}\delta^K_{ij}-\frac{1}{N_{\rm vox}}B_iB_{jk}.
\label{cov3}
\ee
Finally, since we are assuming they have the same signal, the two 1-D histograms will be correlated with each other with covariance
\be
C^{\rm 1D\times 1D}_{ij}=\left<B_{1,i}B_{2,j}\right>-\left<B_{1,i}\right>\left<B_{2,j}\right>=B_{ij}-\frac{1}{N_{\rm vox}}B_{1,i}B_{2,j}.
\label{cov4}
\ee
Derivations of Equations (\ref{cov1}--\ref{cov4}) can be found in Appendix \ref{app:cov}.  With these errors in hand, we define a combined data vector $\vec{D}=(B_{1,j},B_{2,j},B_{ij})$.  The covariance matrix of $\rij$ can then be estimated using the Jacobian matrix $\mathbf{J}=\partial\rij/\partial{\vec{D}}$ via
\be
C_{\mathcal{R}}=\mathbf{J}^TC_D\mathbf{J}.
\label{rcov}
\ee
This implicitly assumes that all of the errors are Gaussian.  We will continue with this assumption throughout this paper, leaving further exploration for future work.

Our error estimation method here has the crucial benefit that we can get an approximate understanding of the error on $\rij$ without specific knowledge of the noise PDF.  After all, a large part of our motivation comes from the difficulty of precisely modeling all of the systematic effects which may enter into the noise PDF, so it would defeat the purpose if we needed to understand all of them for a DDE inference.  With a real data set, we plug the observed $B_{1,j}$, $B_{2,j}$, and $B_{ij}$ into the covariance expressions and get a reasonable estimate of our error bars.

There may be other important effects on the histogram error, for example instrument resolution or large-scale structure effects may cause extra correlation between different bins.  Large-scale structure correlations have been shown to be subdominant to the binomial error, at least in some cases \citep{Polito2022}.  Instrumental effects can generally be minimized by choosing the voxel size to be comparable to or larger than the instrument resolution \citep{Vernstrom2014}.  We leave a more detailed treatment of these possible contributions to future work.

\section{Toy Model}
\label{sec:toy}

As the behavior of the DDE is likely to be quite unintuitive to most, we begin our exploration of its quantitative properties with a simple toy model.  Consider a hypothetical field where our target ``signal" has a Gaussian PDF
\be
\p(\delta)=\frac{1}{\sqrt{2\pi s^2}}\exp\left[-\frac{\delta^2}{2s^2}\right]\equiv\mathcal{N}(\delta,s),
\ee
where our goal is to measure the value of $s$.  Let us also assume that the ``noise" on the observation has a Gaussian PDF $\mathcal{N}(\delta,\sigma_N)$ with width $\sigma_N$, where the exact value of $\sigma_N$ may be unknown due to observational systematics.  In other words, given a map known to contain white noise, can we distinguish any extra variance caused by cosmological emission?  Unless stated otherwise we arbitrarily set $s=\sigma_N=1$ and $N_{\rm vox}=10^3$.

\begin{figure}
\centering
\includegraphics[width=\columnwidth]{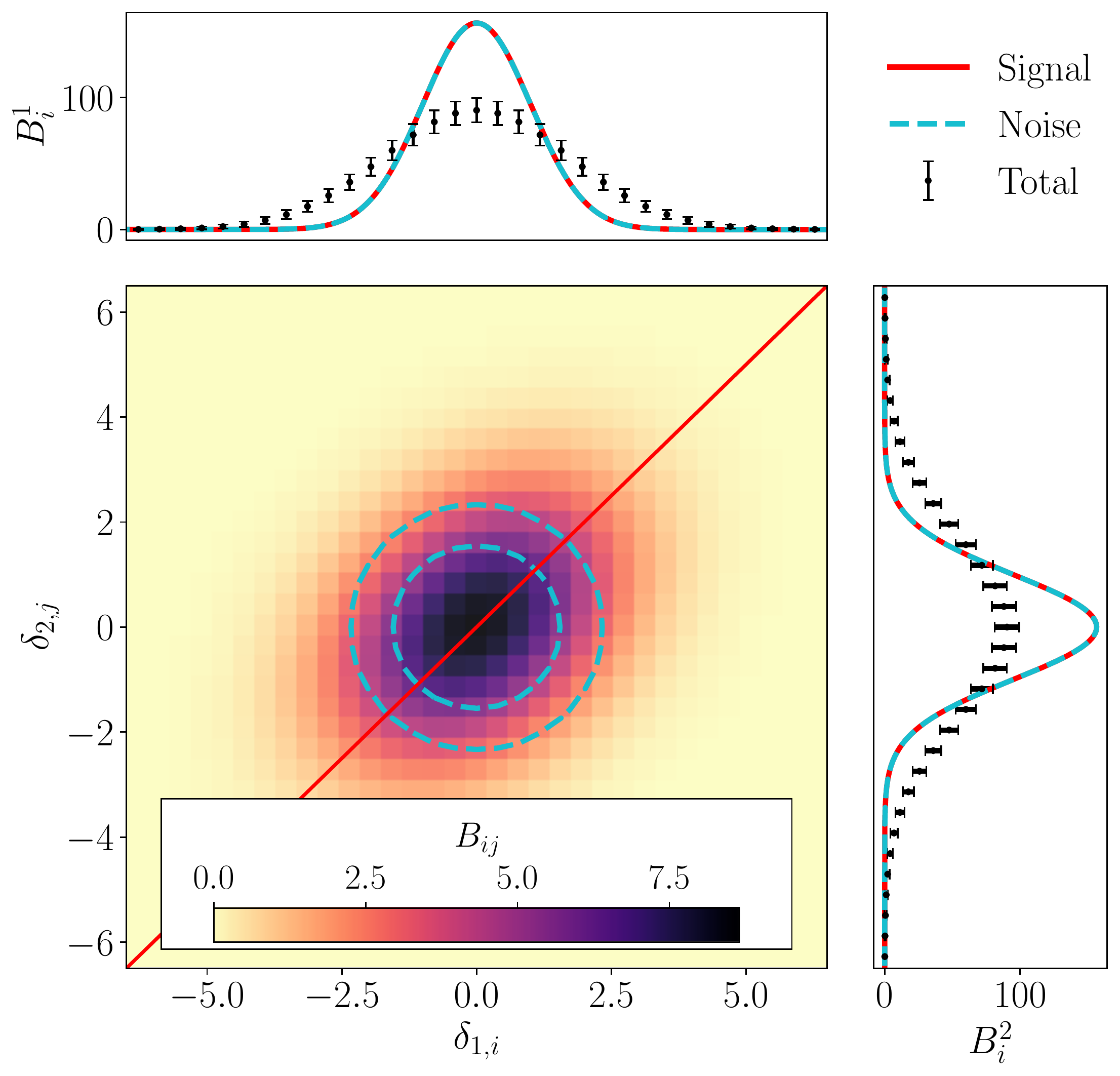}
\caption{(top, right) One-dimensional histograms $B_i^{\rm split}$ for our toy demonstration model, showing the expected signal and noise (red solid and blue dashed, overlapping) as well as the predicted observed histogram with binomial errors (black points and error bars). (Center) Joint two-dimensional histogram of the two data subsets (color bar), showing the one- and two-sigma contours of the signal-only (red) and noise-only (blue) cases.}
\label{fig:toyhist}
\end{figure}

This toy model is obviously not the optimal case for any one-point analysis.  Since $\delta(\bx)$ here is a Gaussian random field, one could obtain all of the relevant information with a standard power spectrum.  However, this model does have the useful feature that it is tractable analytically, allowing us to study our new estimator without getting bogged down in numerical details.

We can imagine measuring this observable with a basic histogram estimator and with our new DDE.  In the histogram case, we expect to obtain
\be
B_i\approx N_{\rm vox}\Delta\delta \mathcal{N}\left(\delta,\sqrt{s^2+\sigma_N^2}\right),
\ee
where we have assumed our bins are narrow enough to approximate the integral in Eq. (\ref{histint}) with its central value.  We can immediately see that $B_i$ is sensitive to a degenerate combination of our target parameter $s$ and the noise $\sigma_N$.  In other words, any error in our prediction of $\sigma_N$ will directly bias $s$.  Figure \ref{fig:toyhist} shows the predicted individual and joint histograms of this toy model split.  It can clearly be seen in the 2D case that the observed histogram is elongated due to the correlation between the two signals.

For the DDE case, we will assume that our ``observation" is split into two parts covering the same signal.  Since each component has only half of the data, we will assign them uncorrelated noise with PDF $\mathcal{N}(\delta,\sigma_N\sqrt{2})$.  The characteristic function of a Gaussian field is also a Gaussian, so for each split we obtain
\be
\widetilde{B}_i^{\rm split}=N_{\rm{vox}}\Delta\delta\exp\left[-\frac{1}{2}\fdel_i^2\left(s^2+2\sigma_N^2\right)\right],
\ee
and for the 2D distribution
\be
\widetilde{B}_{ij}^{\rm split}=N_{\rm{vox}}(\Delta\delta)^2\exp\left[-\frac{1}{2}\left(\fdel_i+\fdel_j\right)^2s^2-\left(\fdel_i^2+\fdel_j^2\right)\sigma_N^2\right].
\ee
Thus, after some algebra, the expected value of the DDE takes the convenient form
\be
\rij=\exp\left[-\fdel_i\fdel_js\right],
\label{toyrij}
\ee
which as expected has no dependence on $\sigma_N$.

\begin{figure}
\centering
\includegraphics[width=\columnwidth]{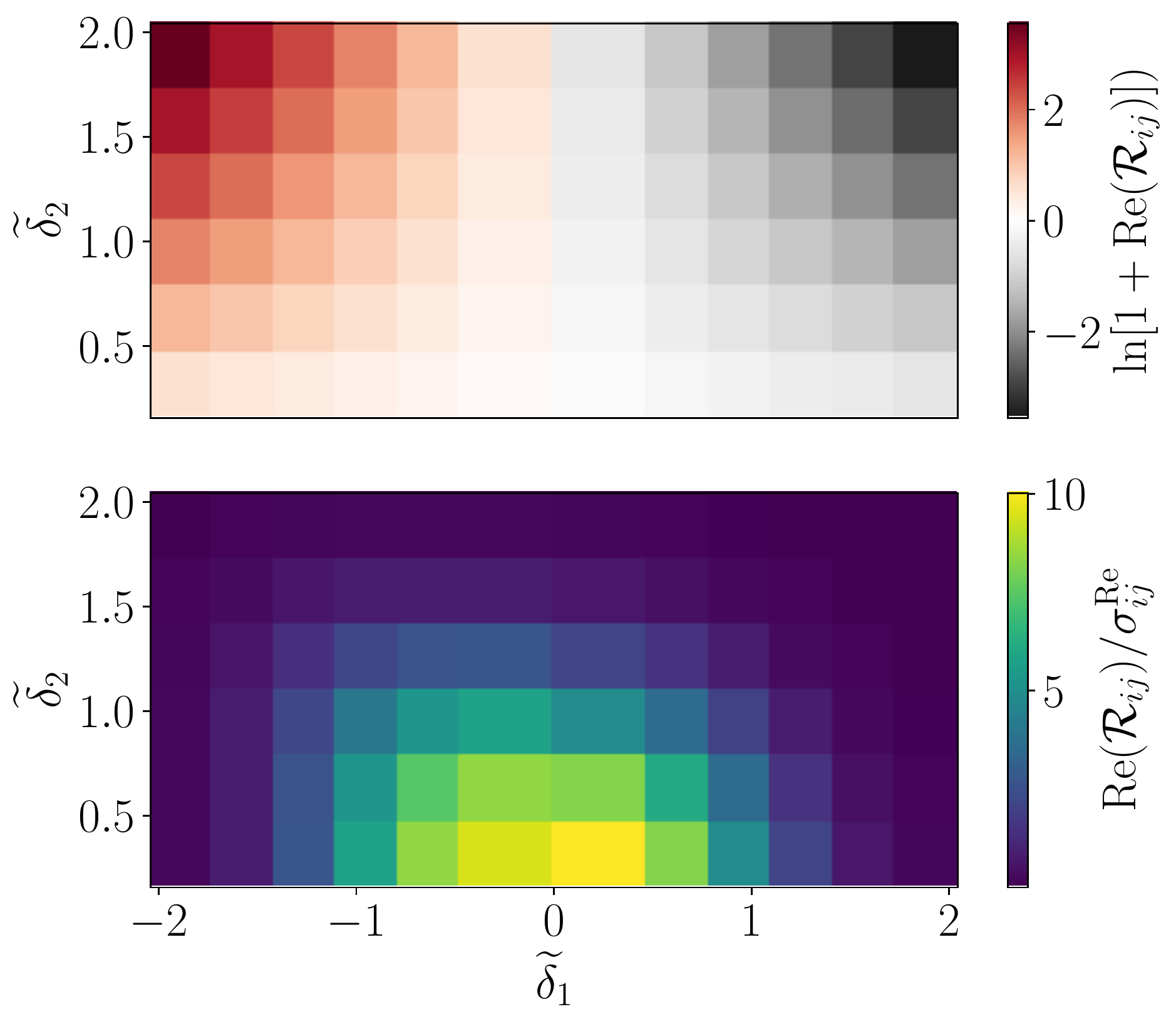}
\caption{(Top) Log of the real part of $\rij$ for our Gaussian toy model. (Bottom) Real part of the toy model $\rij$ normalized by its uncertainty.}
\label{fig:toydde}
\end{figure}

\begin{figure}
\centering
\includegraphics[width=\columnwidth]{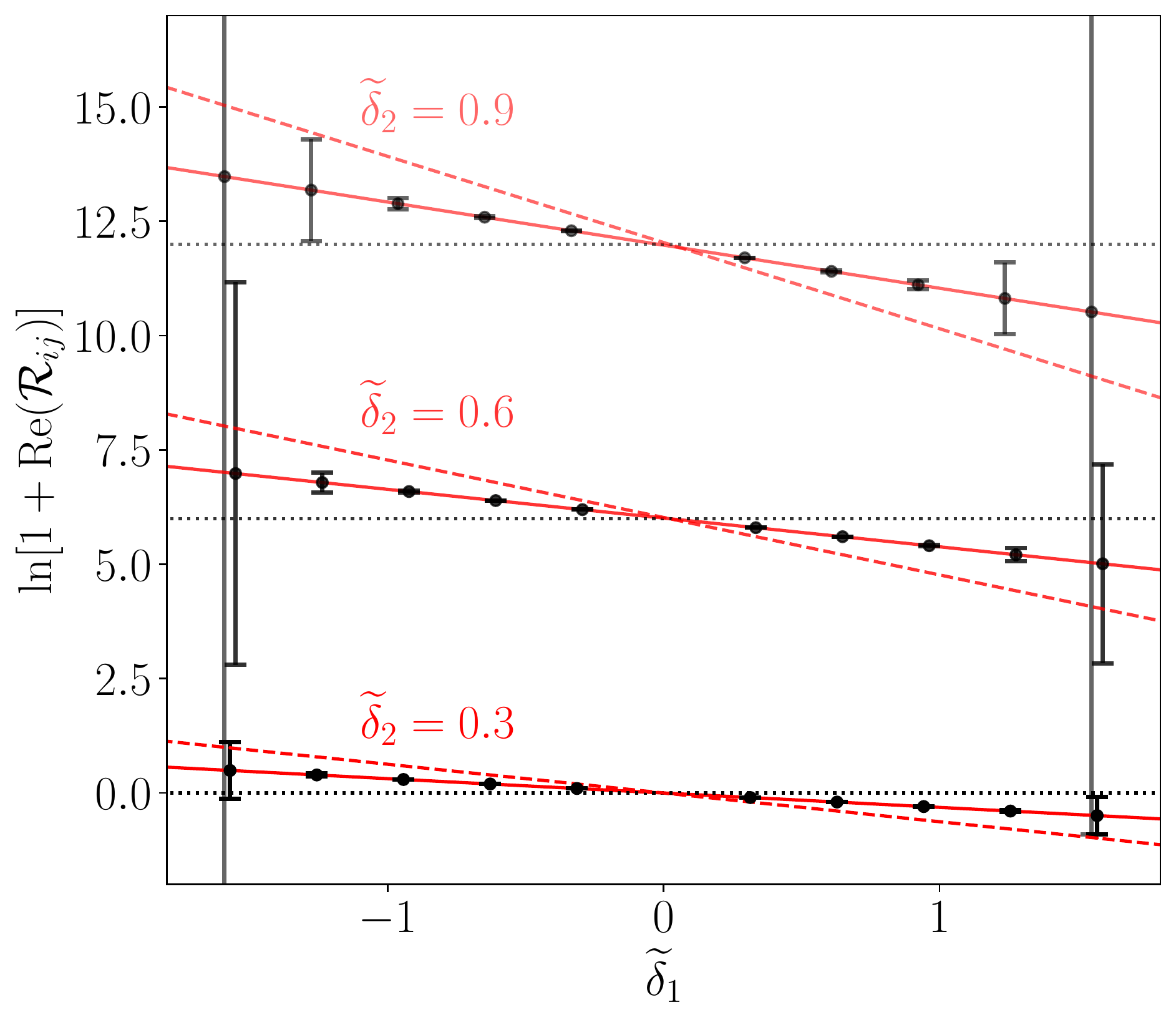}
\caption{Real part of the toy model from Figure \ref{fig:toydde} plotted for individual values of $\fdel_2$, arbitrarily offset vertically.  Red solid lines show the DDE computed only from the toy signal, black points and error bars show the calculation with signal and noise  Black dotted lines show the location of $\rij=0$ for each row.  Red dashed lines show an alternative model where $s=2$ instead of 1.}
\label{fig:toydde_flat}
\end{figure}

\begin{figure}
\centering
\includegraphics[width=\columnwidth]{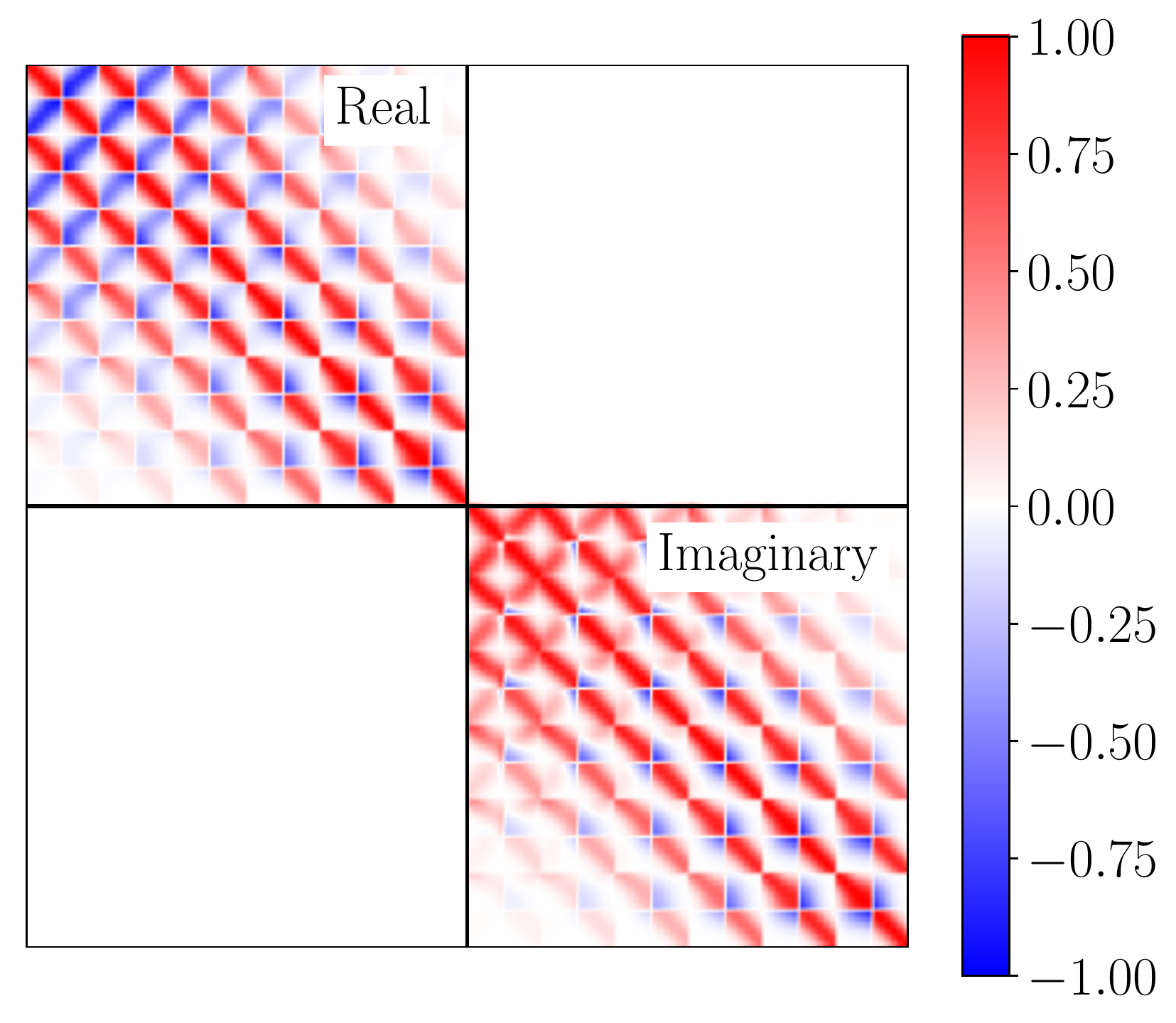}
\caption{Correlation matrix for our toy model DDE with $\rij$ flattened into a one-dimensional vector.  Individual ``x"-shaped features show the correlation for individual rows of $\rij$.  The covariance of the real part appears in the upper left, that of the imaginary part in the lower right, with the other two quadrants showing the (nonexistent for this model) correlation between the two.}
\label{fig:toycov}
\end{figure}

Figure \ref{fig:toydde} shows the real part of $\rij$ predicted by our toy model, with the signal-to-noise ratio plotted in the lower panel.  For readability, individual rows of $\rij$ are plotted in Figure \ref{fig:toydde_flat} along with error bars.  Given the form from Eq. (\ref{toyrij}), we plot the natural log of $\rij$ so that the slope of the lines in Figure \ref{fig:toydde_flat} is proportional to our target parameter $s$.  For comparison, an $\rij$ model with $s=2$ is also plotted in Figure \ref{fig:toydde_flat}.  We compute the imaginary part as well, but it will always be zero in this toy model.  From these plots, we can see an important feature of the DDE.  The highest signal-to-noise is obtained in a finite ``window" at low $\fdel$. This is because the $\widetilde{B}_i$ Fourier transforms in the denominator of $\rij$ will tend towards zero at $\fdel\gtrsim$~a few times $\sigma_N^{-1}$.  This causes the division in $\rij$ to become unstable, blowing up the error bars.  

In general, there are two sources of error on the DDE which will be familiar to those who use power spectra.  First is the instrumental noise contribution, which as just mentioned dominates at high $\fdel$.  Second is a sample variance or cosmic variance contribution from the finite voxel count, which sets the error inside the noise ``window".  There is one key difference with power spectrum errors however.  Figure \ref{fig:toycov} shows the correlation matrix for $\rij$.  Here we can see that the uncertainties on individual $\rij$ bins are extremely correlated.  In our first step of Fourier transforming the data histograms, we mixed the mostly-independent errors in each real space bin across every bin of the characteristic function. Fortunately, we find that as long as we cut off $\fdel$ before the noise becomes too unstable the covariance matrix remains reliably invertible.

\begin{figure}
\centering
\includegraphics[width=\columnwidth]{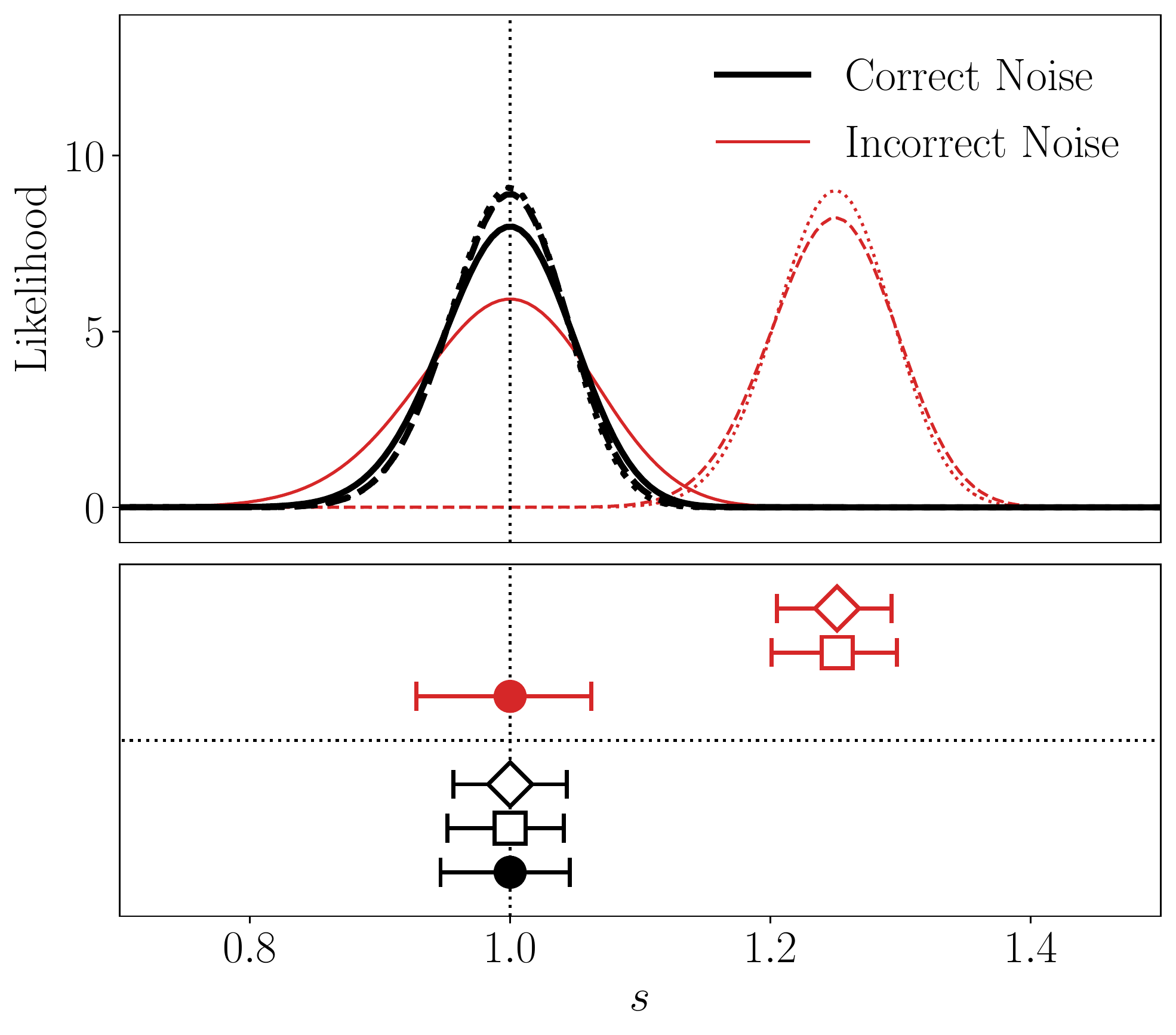}
\caption{(Top panel) Likelihood over our toy model's excess-variance parameter $s$ by fitting the DDE $\rij$ (solid), the 1-D histogram $B_i$ (dashed), and the 2-D histogram $B_{ij}$ (dotted), for the default model (black) and for a model where we assume an incorrect value for $\sigma_N$ (red).  (Bottom) Maximum likelihood and 1-$\sigma$ confidence intervals for $s$ assuming the correct and incorrect $\sigma_N$ values (red and black) for $\rij$ (filled circles), $B_i$ (empty squares), and $B_{ij}$ (empty diamonds).  In both panels, the true $s=1$ value is highlighted.}
\label{fig:toylike}
\end{figure}

Since we have a parameterized model for $\rij$ and its uncertainty, we can use it for inference.  As mentioned above, we currently assume Gaussian errors on $\rij$.  This assumption should be most accurate for the higher signal-to-noise $\fdel$ bins, so we apply a cutoff in $\fdel$ to keep us in the noise ``window" see in Figure \ref{fig:toydde}.  Specifically, since the noise blows up when the error on the denominator of $\rij$ gets large, we make the conservative choice to use only bins where $\fb_i/\sigma(\fb_i)\geq5$.  We estimate the error on $fb_i$ as part of the Jacobian calculation in Eq. (\ref{rcov}).  This should ensure that we avoid the worst non-Gaussianities or instabilities, though our final results are mostly insensitive to this choice due to the large bin-to-bin covariance.  Under these assumptions, we can write down the likelihood for a hypothetical observation of our toy model as
\begin{multline}
\ln\mathcal{L}(s)=-\frac{1}{2}\sum_{ijk\ell}\left(\rij^{\rm obs}-\rij(s)\right)C_{\mathcal{R},ijk\ell}^{-1} \\ \times\left(\mathcal{R}_{k\ell}^{\rm{obs}}-\mathcal{R}_{k\ell}(s)\right),
\end{multline}
up to an overall normalization.  We can write down similar Gaussian likelihoods for inference based on the 1-D and 2-D histograms as well.  Since we only have one parameter $s$ and the model is fast to evaluate, we can compute the likelihood directly without resorting to Monte Carlo or Fisher methods.  Figure \ref{fig:toylike} shows the result.  For the default model, we can see that fits to $B_i$, $B_{ij}$, and $\rij$ all recover the true $s=1$.  The DDE estimator performs the worst of the three, because we have thrown away some information in the calculation of $\rij$.

However, this calculation assumes that we have perfect understanding of our imaginary ''instrument" and any systematics which it adds to our data.  In real life this is often not true, so we plot likelihoods for an additional case where we believe our noise to have $\sigma_N=1$ when in the actual data it is higher by $25\%$ due to some unforseen complication.  In this case, we see that the histogram estimates are biased, attempting to fit the excess noise by increasing $s$.  The DDE, on the other hand, still prefers the true value.  This is a quantitative illustration of our statement that the DDE is unbiased by uncorrelated systematics.  The fitted error on $s$ does still increase, as the added noise propagates through to larger errors on $\rij$.

Obviously, for this example, one could easily perform a joint fit for $s$ and $\sigma_N$ and marginalize over the noise.  The DDE will remain unbiased, though, for any other uncorrelated error which may appear in the data as well.  In a real, messy experiment it may not be possible to precisely model every possible systematic.  We thus have an example of the common bias-vs.-variance trade-off in data analysis.  If we are confident that we understand all aspects of a data set, the histogram estimators $B_i$ and $B_{ij}$ will always be more informative.  But if we suspect that the histograms may be biased, we can cancel out all of that bias using $\rij$ at the cost of a modest amount of overall sensitivity.

\section{Intensity Mapping Model}
\label{sec:lim}

\begin{figure}
\centering
\includegraphics[width=\columnwidth]{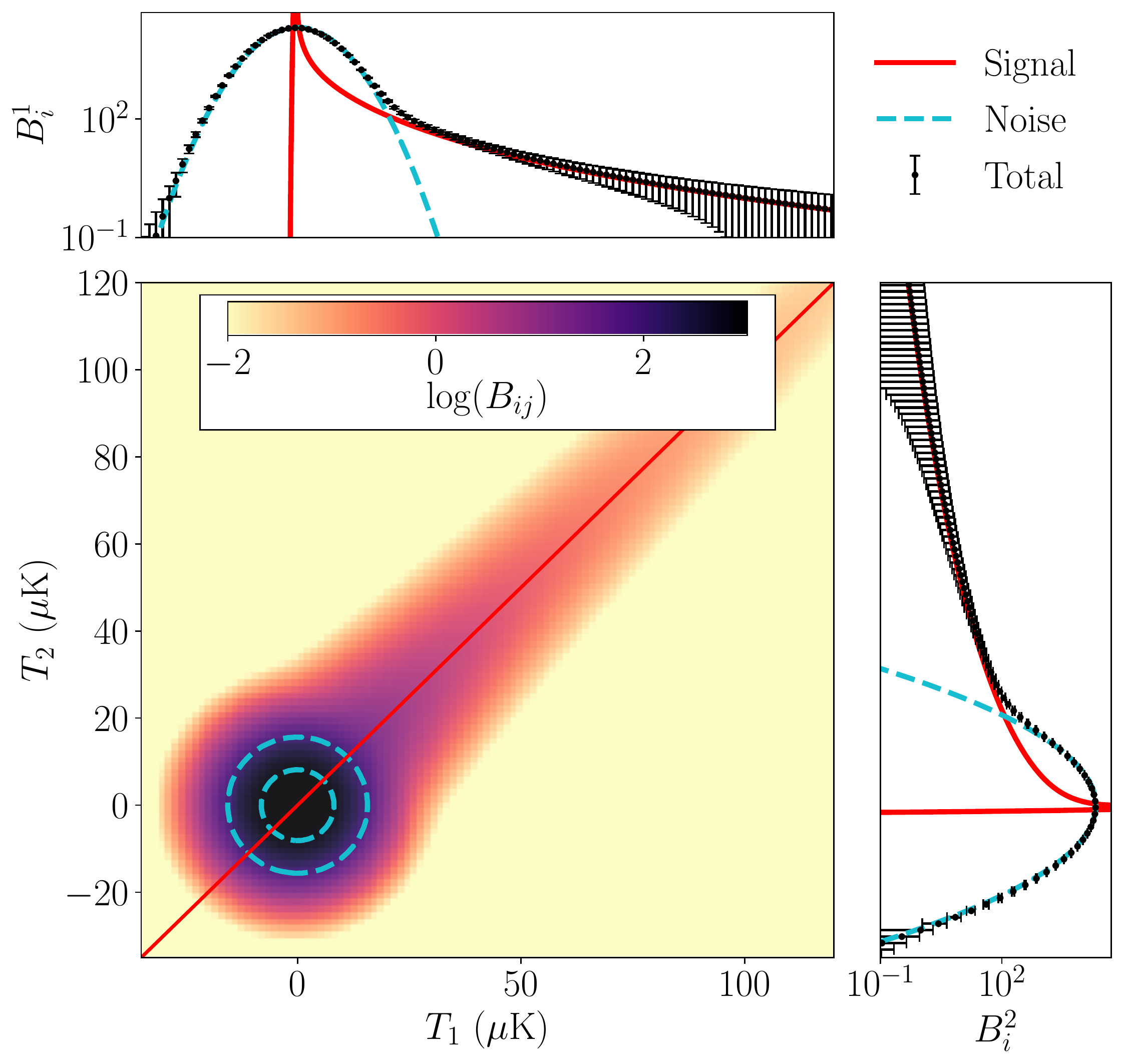}
\caption{Same as Figure \ref{fig:toyhist} for our COMAP-EoR model, showing the 1-D histograms (top and right) and the 2-D histogram (center).  Histograms are now plotted on a log scale as the LIM model has a larger dynamic range.}
\label{fig:limhist}
\end{figure}

Now we will demonstrate how the DDE behaves under a more physical model, to show that it is indeed useful for a realistic non-Gaussian field.  Specifically, we will forecast the DDE's effectiveness for a near-future line intensity mapping (LIM) \citep{Kovetz2017,Bernal2022}.  LIM surveys are observations of the intensity fluctuations in a target emission line over large cosmological volumes performed with relatively coarse spatial resolution.  Many surveys target species like CO or \ion{C}{ii} which are sourced within star-forming galaxies, so in practice they observe the aggregate emission from all of the galaxies in a single resolution element \citep{Keating2016,Yang2019,Cataldo2021,CCAT2021,Sun2021,Anderson2022,Bethermin2022,esI,Karkare2022}.  By targeting narrow emission lines, LIM can obtain high redshift resolution by observing in many closely-spaced frequency bands.  

LIM data are a particularly exciting use-case for one-point statistics.  Because galaxies are extremely complex, nonlinear systems, the resulting intensity field will in general be highly non-Gaussian.  That non-Gaussianity is primarily sourced within what are effectively point sources,  as opposed to larger-scale mode couplings which are more easily studied with bispectra or similar.  The histogram $B_i$ of a line intensity map is referred to as the Voxel Intensity Distribution (VID).  \citet{Breysse2017} showed using probability of deflection, or $\p(D)$, analysis \citep{Scheuer1957} that the VID enables direct measurement of the line luminosity function of the unresolved galaxy population, which can be connected to many interesting physical quantities \citep{Breysse2016,Ihle2019,Bernal2021,Bernal2021a,Chung2021,Libanore2022,Pullen2022}.

In a LIM survey, our primary observable will be the total intensity $T(\mathbf{x})$ in a voxel at location $\mathbf{x}$.  We write intensity here as a brightness temperature as is the convention for lower-frequency LIM surveys like our below demonstration.  Assuming our line is sourced within galaxies which are small compared to the voxel volume $V_{\rm vox}$, we have
\be
T(\mathbf{x})=\frac{C_{LT}}{V_{\rm{vox}}}\sum_{j=1}^{N_{\rm gal}}L_j,
\ee
where voxel $i$ contains $N_{\rm gal}$ galaxies the $j$'th of which has line luminosity $L_j$.  For a survey reporting brightness temperature, the conversion factor is
\be
X_{LT}\equiv\frac{c^3(1+z)^2}{8\pi k_B\nu_{\rm em}^3 H(z)},
\ee
\citep{Lidz2011} where the line is emitted at redshift $z$ with rest frequency $\nu_{\rm em}$, $c$ is the speed of light, $k_B$ is Boltzmann's constant, and $H(z)$ is the Hubble parameter.  Typically, we model the line luminosity by assuming some mean relationship $L(M)$ between a galaxy's luminosity and its host halo mass $M$, often with some scatter about that relation to account for other important properties \citep[see, e.g.][]{ Lidz2011,Pullen2013,Li2016,Yang2022}. The most up-to-date analytic formalism for predicting $B_i$ from $L(M)$ was presented in \citet{Breysse2022bias}, based on similar calculations for the CIB and weak lensing \citep{Thiele2019,Thiele2020}.  This derivation is reproduced in Appendix \ref{app:PofD} for the convenience of the reader.

\begin{figure*}
\centering
\includegraphics[width=\textwidth]{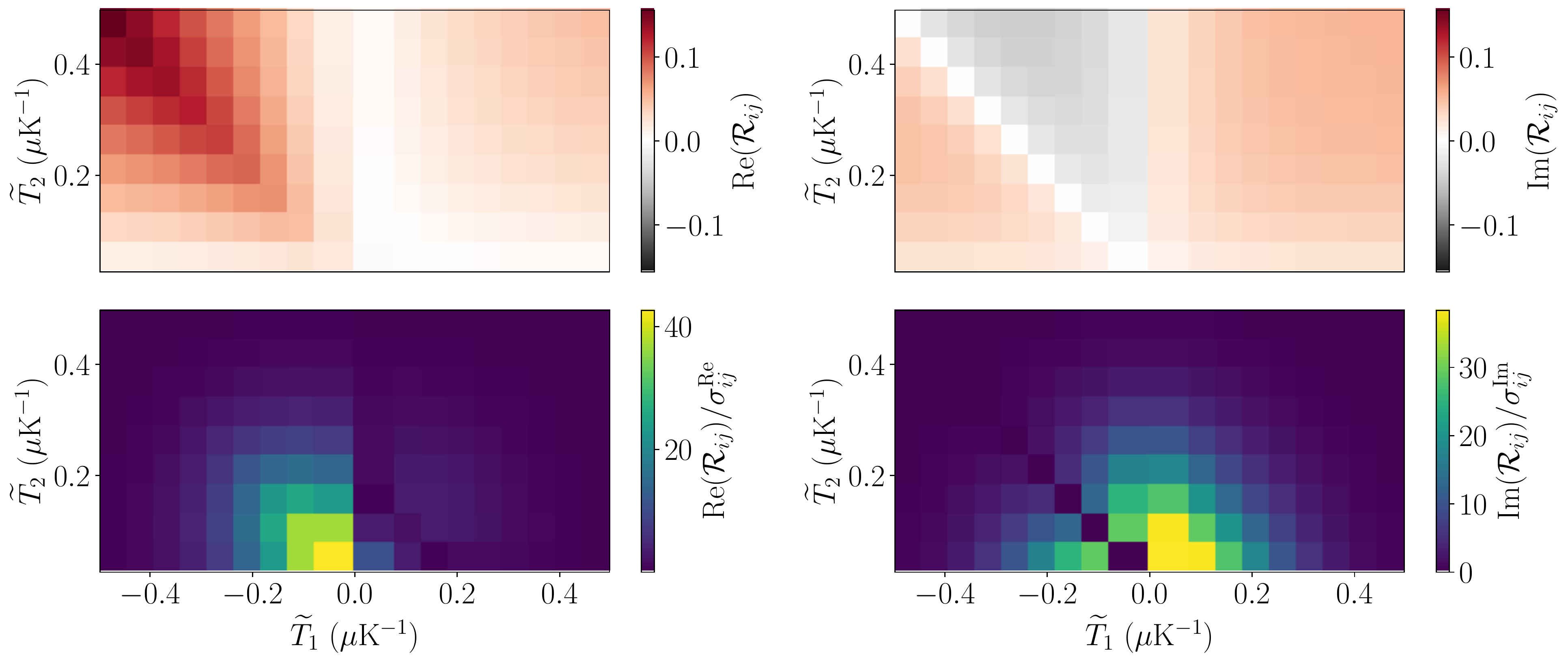}
\caption{Same as Figure \ref{fig:toydde}, but for the LIM model, now showing the real part (left) and the imaginary part (right) of $\rij$ and their respective signal-to-noise.}
\label{fig:limdde}
\end{figure*}

For our example experiment, we choose the Epoch of Reionization expansion of the Carbon Monoxide Mapping Array Project \citep[COMAP-EoR,][]{esVII}.  COMAP-EoR, building on the existing COMAP Pathfinder \citep{esI}, seeks to map the rotational transitions of CO molecules in several bands over a broad redshift range.  For maximum clarity we want a fairly high signal-to-noise example, so we choose to demonstrate the DDE forecasting the single lowest redshift bin $z=2.4-2.8$, corresponding to the 115 GHz CO(1-0) line observed at $\nu_{\rm obs}=26-30$ GHz.  We assume the map has white noise at levels predicted from the specifications in \citet{esVII}.  We do not attempt to analytically apply an instrument transfer function to the VID, leaving that for future work.  We attempt to account for the most important aspect, though, by subtracting the overall mean of the signal from our hypothetical maps as discussed in \citet{Breysse2017}.  We assign a voxel to be a rectangular area defined by the beam full-width half-maximum in the plane of the sky and by a single frequency channel along the line of sight.  For the $L(M)$ model, we use the double power law parameterization from \citet{Padmanabhan2018}
\be
L(M)=\frac{C}{\left(M/M_*\right)^A+\left(M/M_*\right)^B},
\label{LofM}
\ee
with maximum likelihood values for the free parameters $p_\alpha\equiv(A,B,C,M_*,\sigma_{\rm sc})$ given by the ``UM+COLDz+COPSS" model from \citet{esV}.

\begin{figure*}
\centering
\includegraphics[width=\textwidth]{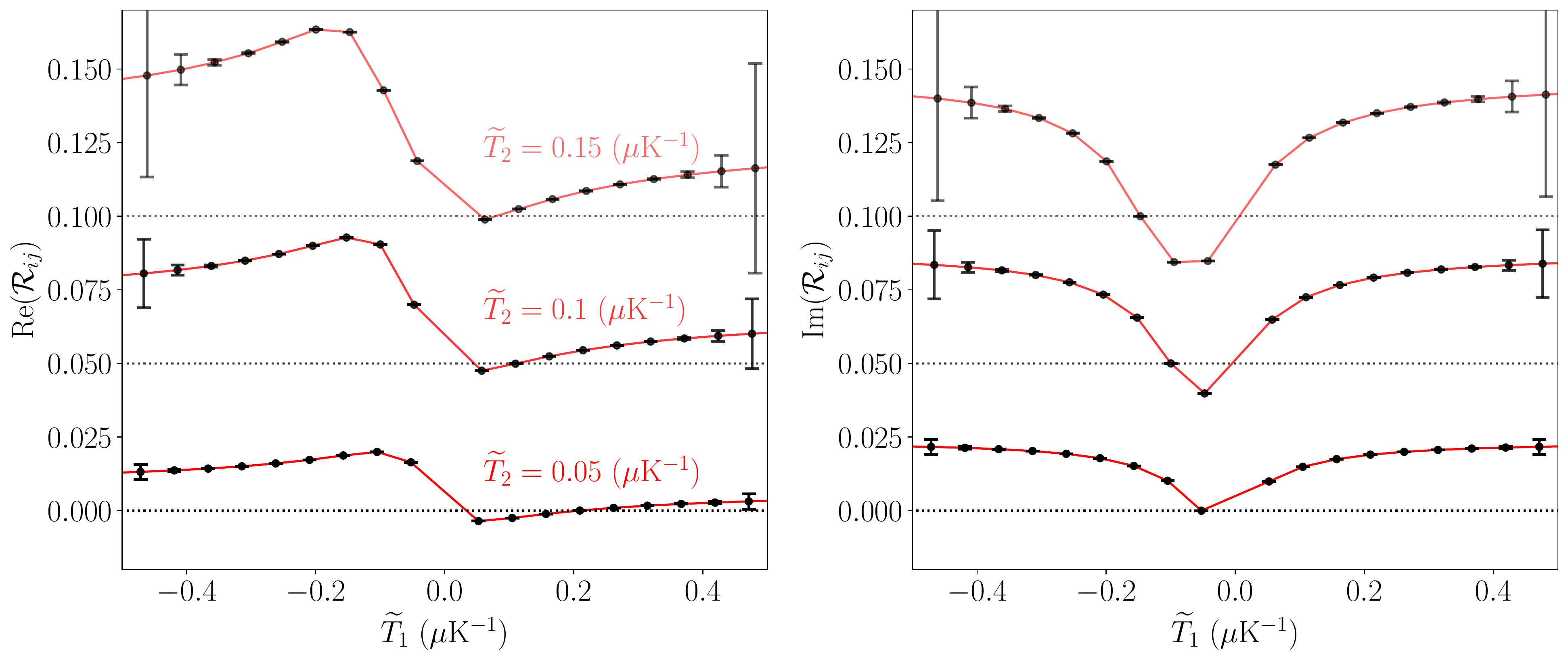}
\caption{Same as Figure \ref{fig:toydde_flat}, showing individual rows of the LIM model DDE, again showing the real part (left) and the imaginary part (right)}
\label{fig:limdde_flat}
\end{figure*}

Figure \ref{fig:limhist} shows the 1- and 2-D histograms obtained from this model and experiment.  In this model, which is very high signal-to-noise, we see the peak of the Gaussian noise PDF and the long tail of the line luminosity function.  The tail is likely more dramatic than it would be in reality, as we have not accounted for the loss of small-scale information in the instrument transfer function \citep{esIV}.  Though the DDE lacks the convenient analytic form it had in the toy model, it is still straightforward to Fourier transform these histograms to produce the DDE shown in Figures \ref{fig:limdde} and \ref{fig:limdde_flat}.  The shape of $\rij$ is more complicated, but we see the same windowing effect wherein the error bars grow rapidly at large $\ft$.  Because our original PDFs are no longer symmetric around $T=0$, we also have a nonzero imaginary part to $\rij$.  The imaginary part displays the same windowing behavior, we also see a noticeable feature where ${\rm Im}(\rij)=0$ along the line $\ft_1=-\ft_2$.  This is an artifact of our choice to make the two signals identical.  Examining Equation (\ref{Rsignal}), we see that along this line we have 
\be
\ftp^S_{\rm 2D}(\ft_1,\ft_2)=\ftp^S_{1D}(\ft_1+\ft_2)=1,
\ee
since the PDFs are required to be normalized.  Meanwhile, because the PDFs are real, symmetry conditions will cancel out the imaginary parts in the denominator of $\rij$, leaving no imaginary part\footnote{Unlike the $\ft=0$ bins, $\rij$ when $\ft_1=-\ft_2$ is merely expected to be zero, rather than required.  $\ft=0$ bins would have zero error bar, which means we cannot include them in our covariance matices. The $\ft_1=-\ft_2$ bins have nonzero error and are thus safe to include.}.

\begin{figure}
\centering
\includegraphics[width=\columnwidth]{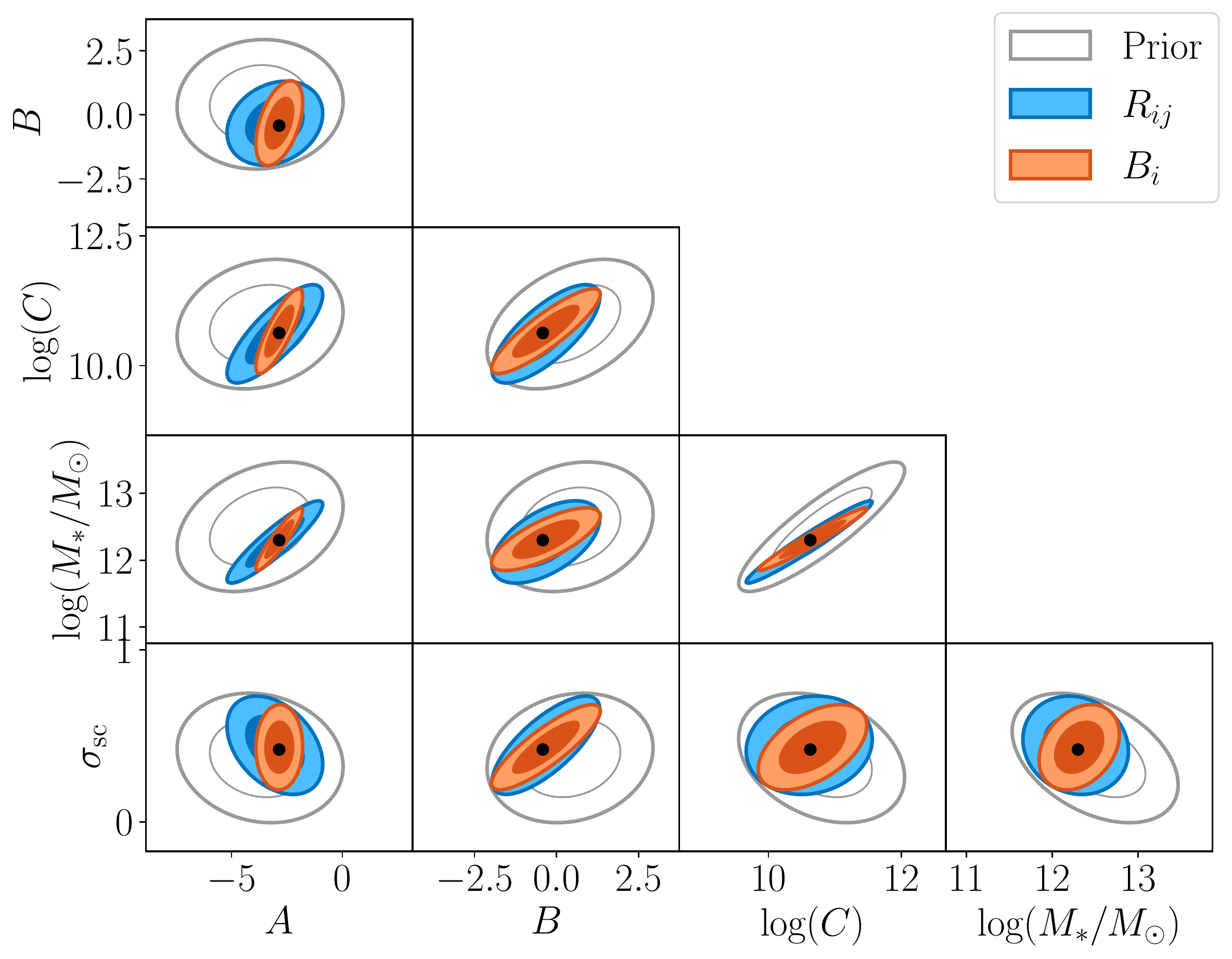}
\caption{Fisher forecasts for the CO $L(M)$ parameters using the Gaussianized priors from \citet{esV} (gray empty), the VID $B_i$ (red), and the DDE $\rij$ (blue).  Light and dark (thick and thin) ellipses show the 68 and 95\% confidence intervals.  Black dots show the true input values of the parameters.  Note that the prior ellipses are not centered at the true input values, as the Gaussianization process causes an offset from the peak of the true prior distribution.}
\label{fig:lim_fisher}
\end{figure}

Because we a model for the the underlying $L(M)$, we can see how the LIM DDE varies when we change the model parameters, which in turn will let us constrain those parameters from a measurement.  Variations of $\rij$ with the model parameters are shown in Appendix \ref{app:variation}.  We can estimate the constraining power of this DDE measurement using a Fisher analysis \citep{Fisher1935}.  The Fisher matrix, which gives the covariance matrix of the model parameters at their maximum likelihood values, is given by
\be
F_{\alpha\beta}=\sum_{ijk\ell}\frac{\partial\rij}{\partial p_{\alpha}}C^{-1}_{\mathcal{R},ijk\ell}\frac{\partial\mathcal{R}_{k\ell}}{\partial p_{\beta}},
\ee
and similarly for $B_i$.  Because the Fisher calculation approximates the likelihood as Gaussian, we will use the Gaussianized priors on our model parameters $p_{\alpha}$ from Table 5 of \citet{esV}.  Figure \ref{fig:lim_fisher} shows the results of the Fisher forecast.  We can clearly see that, though the DDE is slightly less constraining than the VID, it retains most of the information in the data.

\begin{figure}
\centering
\includegraphics[width=\columnwidth]{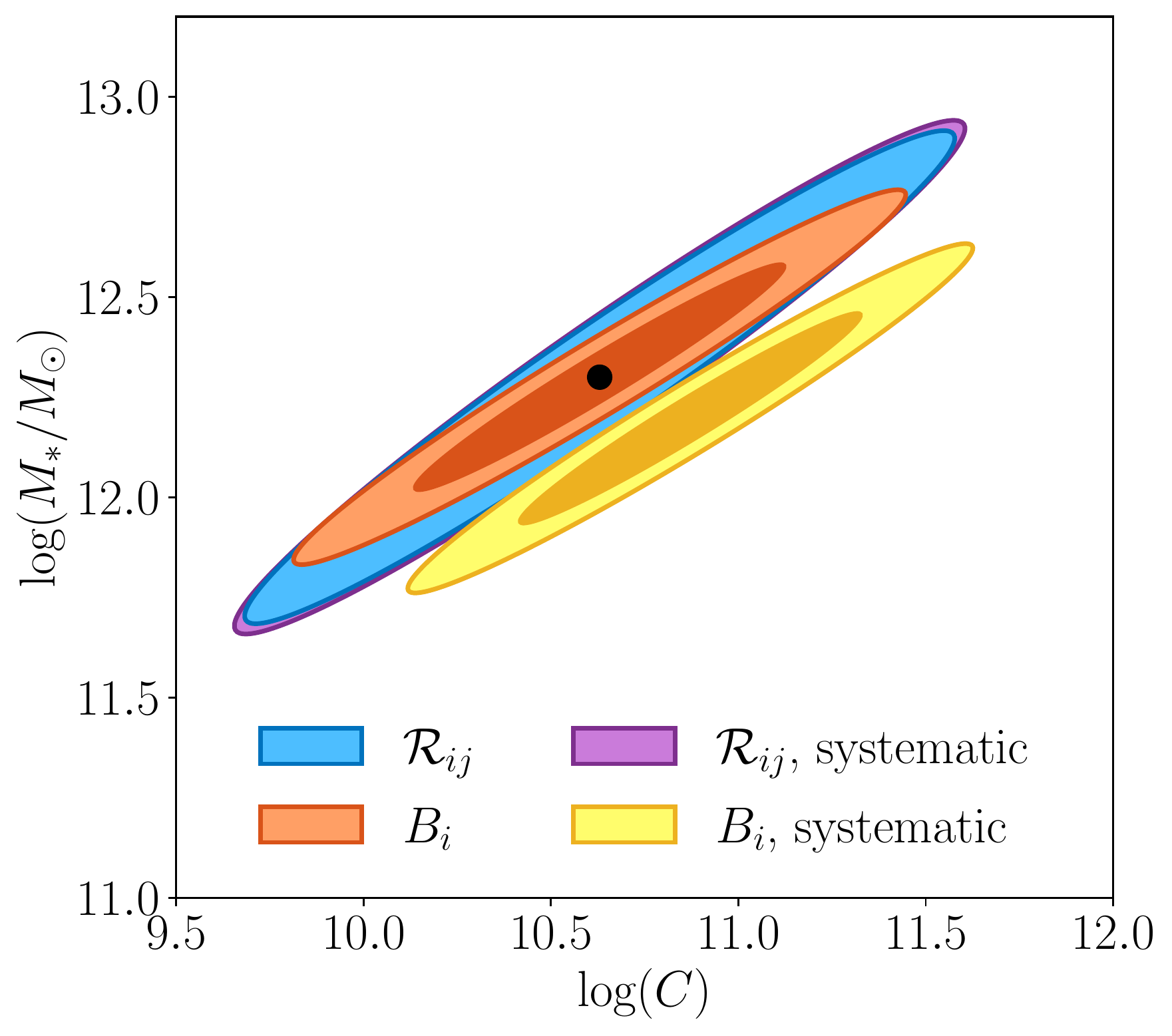}
\caption{Component of the Fisher forecast for the amplitude $C$ and knee $M_*$ model parameters showing the effect of adding an uncorrelated systematic to our LIM data which mimics a brighter version of the CO model.  Red and blue ellipses show the $B_i$ and $\rij$ results from Figure \ref{fig:lim_fisher}, yellow and purple ellipses show the effect on $B_i$ and $\rij$ (respectively) of adding the new systematic.}
\label{fig:lim_fisher_bias}
\end{figure}

Our final illustration will be to demonstrate once more that $\rij$ is unbiased by uncorrelated systematics.  There are virtually infinitely many different systematics which could come into a LIM observation, so rather than attempt to model any specific effect we will simulate something of a worst-case scenario.  We will add to each mock observation an additional component with exactly the same PDF shape as our CO model.  In order to maximize the contrast, we will make this excess noise three times brighter than the true CO model (i.e., we will increase the ``$C$" parameter by a factor of three).  We will still treat this interloper as an uncorrelated systematic, so though we add it to both halves of our data we assume the individual voxel values are still uncorrelated.  Figure \ref{fig:lim_fisher_bias} shows how adding this systematic affects the Fisher forecast.  For readabilty, we show only one set of parameters, but similar effects appear in other parts of parameter space.  We can clearly see that, with the systematic included, the $B_i$ forecast is biased away from the true value, but the $\rij$ fit is still in the right place, if with slightly higher uncertainty.

\begin{figure}
\centering
\includegraphics[width=\columnwidth]{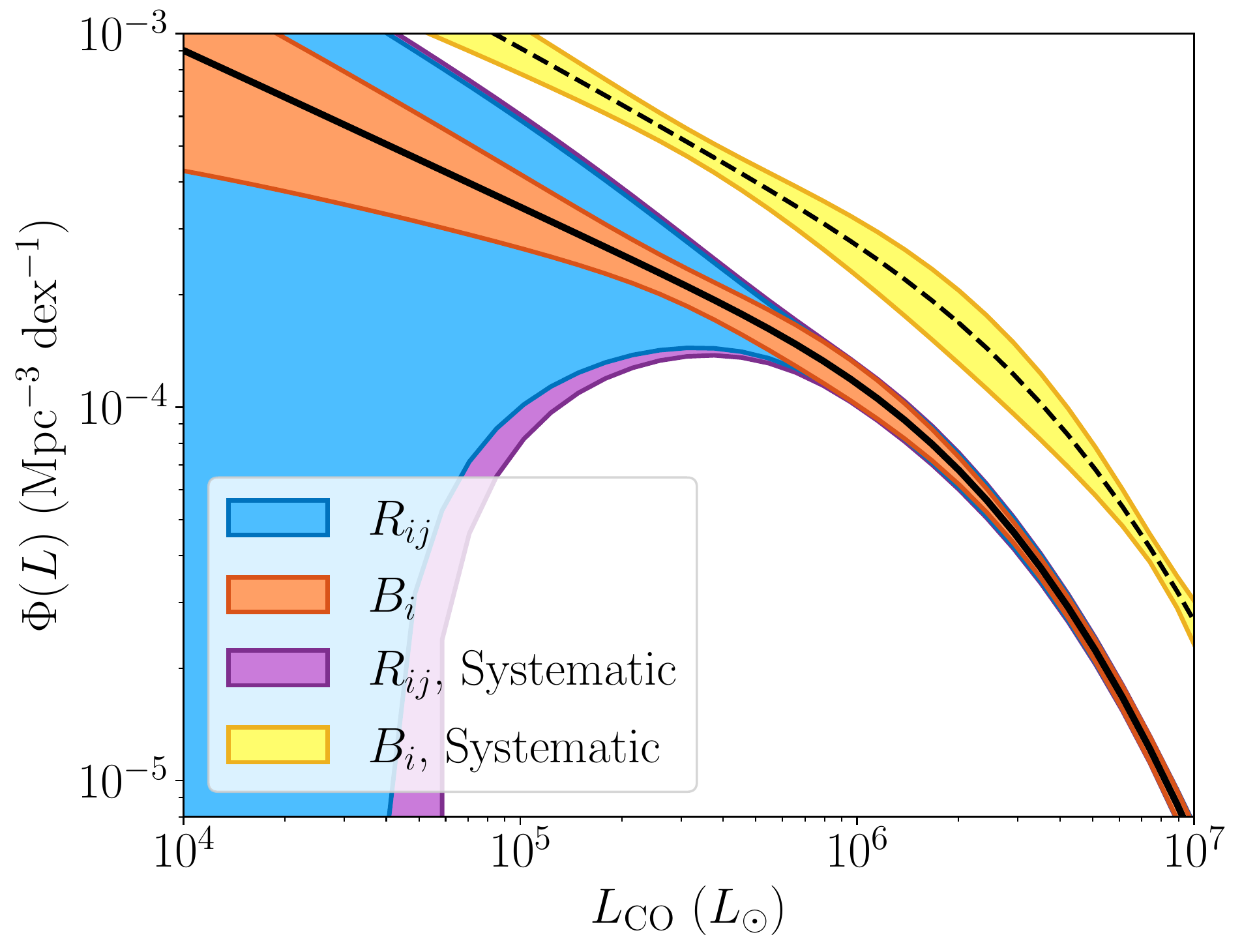}
\caption{Constraints on the CO luminosity function $\Phi(L)$ (black) from our hypothetical COMAP-EoR observation.  Red bands show the 1-$\sigma$ confidence interval for the fiducial $B_i$ measurement, blue show the fiducial $\rij$ measurement.  Yellow and purple bands show the effect of adding the systematic from Figure \ref{fig:lim_fisher_bias} on the $B_i$ and $\rij$ results.  The maximum likelihood systematics-contaminated $B_i$ fit is shown as a black dashed line.}
\label{fig:dndL}
\end{figure}

Figure \ref{fig:dndL} summarizes the results of our Fisher calculations into errors on the CO luminosity function $\Phi(L)$.  This is the quantity which would actually be integrated to obtain the cosmic star formation rate \citep{Breysse2016} or the molecular gas abundance \citep{Keating2020,esI}.  We show fits for the 1-D VID and the DDE both with and without a signal-mimicking systematic.  As before, $B_i$ outperforms $\rij$ in terms of precision in both cases.  In particular, the DDE is much less informative about the faint end of the luminosity function.  However, we also see again that, because it has no way to distinguish a signal on the sky from an observational systematic, the VID becomes significantly biased when adding a systematic.  The 1-D histogram is forced to model the excess emission as signal, resulting in the brighter fit.  Thus we see the same qualitative bias-variance tradeoff in our detailed LIM model that originally appeared in the toy model.

\section{Discussion}
\label{sec:discussion}

The previous two examples show that our new DDE estimator meets the criteria for a cross-correlation analogue.  The two models shown here are quite simplified for illustration purposes, but the lessons should transfer to more sophisticated, realistic calculations.  The results should also generalize to any other pair of one-point data sets, as the basic principles only care about the statistics of the fields and not any underlying physics.

We have not explored here in detail the utility of the DDE in the case where the two signal fields are merely correlated rather than identical.  The modeling in this case would be slightly more complicated, as one would need to model both the individual fields and their correlation.  For a LIM cross-correlation, this could be done by assuming an individual $L(M)$ model for each line and a correlated scatter between them \citep{Schaan2021,Yang2022}.  A detailed case study on this class of problem can be found in \citet{Chung2022}.

The DDE shares many qualities in common with the CVID estimator from \citet{Breysse2019}.  Both use the fact that characteristic function of an observable is the product of those of the signal and noise to cancel out an undesired noise component.  The CVID cases differs in that it assumes one of the fields being cross-correlated is essentially a binary variable (a spectroscopic galaxy survey in the case of \citet{Breysse2019}).  If a field can only take two values it is obviously hard to Fourier transform its histogram.  We leave more general exploration of the relationship between these to estimators for future work, save to note that there may be some connection to the optimal observables concept discussed in \citet{Cheng2019}.

Since this is a first theoretical demonstration of the DDE concept, we have left out many important effects which come up in real data.  For some effects, it is easy to show that they will not alter the basic utility of the DDE.  As an example, often the noise in a map will not be homogeneous, but will have a different value at every point.  This by itself will not affect the DDE, but it is common practice to weight the observed map by the expected noise level to obtain a more optimal estimate of the signal \citep{esIV,Hivon2002}.  The resulting ``pseudo-power spectra" must be carefully modeled, but provide better signal-to-noise than their unweighted equivalents.  For the DDE, one could perform a similar procedure, modeling the PDF of the weighted intensity.  As long as the $\rij$ were modified correctly, this should not induce any extra correlations between the maps and the DDE should still work.  On the other hand, any systematics which appear correlated between the two data sets will not drop out of the DDE and must be treated separately.

It may perhaps be slightly worrying that the LIM luminosity function constraints from the DDE fare so much worse at low luminosity than the $B_i$ ones.  This is especially true since detecting faint galaxies is a primary science goal of LIM.  However, as stated above, this sacrifice may be worthwhile when attempting to reject biasing contaminants from a data set.  For example, \citet{Chung2022} show that correlating two line maps using the DDE can separate out emission at a target redshift from interloper lines in the same frequency band.  The loss will be mitigated somewhat when a DDE fit is combined with a standard power spectrum, as the power spectrum will constrain overall integrals of the distribution \citep{Ihle2019}.  This will require a model for the covariance between the power spectrum and $\rij$, but this should be obtainable using either the simulation method from \citet{Ihle2019} or the analytic formalism of \citet{Polito2022}.

More generally, though, a valuable topic for future work will be to study the optimality of the DDE as defined here.  As the name implies, $\rij$ is at its heart a deconvolution, and those are known to be unstable when the ``filter" appearing in the denominator approaches zero.  This does not cause any great issue for our analysis, since as discussed the error bars expand in sync with the instability.  It may be possible though to write down a more optimal deconvolution, perhaps involving some type of Wiener filtering \citep{Zaroubi1995}, which retains more of the high-$\fdel$ information.

\section{Conclusion}
\label{sec:conclusion}

We have demonstrated a new, characteristic-function based estimator which transfers the useful aspects of familiar two-point cross-correlations to the measurement of one-point statistics.  The Deconvolved Distribution Estimator $\rij$, computed by deconvolving the joint PDF of a pair of density fields by their individual PDFs, cancels out any noise or systematic which is uncorrelated between the two while retaining nearly all of their one-point information.  

We showed two example correlations here, but the possible use case of the DDE is nearly as broad as the cross-power spectrum.  Any pair of correlated, non-Gaussian fields will contain information in their joint PDF, which can be accessed by the DDE. For example, correlating galaxy positions with weak lensing maps could probe non-Gaussian aspects of the galaxy-dark matter connection.  Correlating those same galaxies with Compton-y maps could better connect physical properties of clusters to the Sunyaev-Zeldovich effect. Care will need to be taken to accurately apply this new technique to unfamiliar types of data, but the DDE has the potential to shed new light on a wide range of past and future cosmological observations.

\section*{Acknowledgements}

The authors would like to thank Christopher Anderson, Jos\'e Bernal, Yun-Ting Cheng, Kieran Cleary, Adam Lidz, Anthony Pullen, Gabriela Sato-Polito, Eric Switzer, and the members of the COMAP and EXCLAIM collaborations for useful conversations.  PCB was supported by the James Arthur Postdoctoral Fellowship. DTC is supported by a CITA/Dunlap Institute postdoctoral fellowship. The Dunlap Institute is funded through an endowment established by the David Dunlap family and the University of Toronto. The University of Toronto operates on the traditional land of the Huron-Wendat, the Seneca, and most recently, the Mississaugas of the Credit River; DTC and others at the University of Toronto are grateful to have the opportunity to work on this land. DTC also acknowledges support through the Vincent and Beatrice Tremaine Postdoctoral Fellowship at CITA.

This research made use of Astropy,\footnote{http://www.astropy.org} a community-developed core Python package for astronomy \citep{Astropy2013,Astropy2018}, as well as the NumPy \citep{Numpy2020} and SciPy \citep{Scipy2020} packages.

\section*{Data Availability}

The data underlying this article will be shared on reasonable request to the corresponding author.



\bibliographystyle{mnras}
\bibliography{references} 




\appendix

\section{Histogram Covariances}
\label{app:cov}

Here we derive the covariances quoted in Equations (\ref{cov1}--\ref{cov4}).

\subsection{1-D Histogram}
First, for clarity, we derive the standard form of the multinomial covariance matrix.  Introduce an indicator variable $\di{\alpha}{i}$, defined that $\di{\alpha}{i}=1$ if the value in voxel $\alpha$ falls into histogram bin $i$.  This implies that
\be
B_i=\sum_{\alpha=1}^{N_{\rm vox}}\di{\alpha}{i},
\ee
Or equivalently, since all voxels are independent, $\langle\di{\alpha}{i}\rangle=B_i/N_{\rm{vox}}$.  This gives the useful property that $(\di{\alpha}{i})^2=\di{\alpha}{i}$.

With this in mind, we can expand the first term of the covariance as
\be
\left<B_iB_j\right>=\left<\left(\sum_{\alpha}\di{\alpha}{i}\right)\left(\sum_{\beta}\di{\beta}{j}\right)\right>=\sum_{\alpha,\beta}\left<\di{\alpha}{i}\di{\beta}{j}\right>.
\ee
We can separate this expectation into two terms, one where $\alpha=\beta$ and one where the two are different:
\be
\left<B_iB_j\right>=\sum_{\alpha}\left<\di{\alpha}{i}\di{\alpha}{j}\right>+\sum_{\alpha\neq\beta}\left<\di{\alpha}{i}\right>\left<\di{\beta}{j}\right>.
\ee
The first term is zero unless $i=j$, since one voxel cannot fall into two different bins.  Carrying out the sums gives us
\be
\left<B_iB_j\right>=B_i\delta^K_{ij}+\left(1-\frac{1}{N_{\rm vox}}\right)B_iB_j.
\ee
Adding this into the full covariance expression then yields
\be
C_{ij}^{\rm 1D}=B_i\delta_{ij}^K-\frac{1}{N_{\rm vox}}B_iB_j.
\ee
An identical procedure can be carried out for the 2-D histogram to produce Eq. (\ref{cov2}).

\subsection{Correlation Between 1-D and 2-D Histograms}
Because they are drawing from the same underlying data, we expect correlations in measurements of $B_i$ and $B_{ij}$.  Specifically, the definition of the two requires that $B_i=\sum_jB_{ij}$.

We again write the first term of the covariance, now using indicator variables for both the 1-D and 2-D histograms
\be
\left<B_iB_{jk}\right>=\left<\left(\sum_{\alpha}\di{\alpha}{i}\right)\left(\sum_{\beta}\di{\beta}{jk}\right)\right>=\sum_{\alpha,\beta}\left<\di{\alpha}{i}\di{\beta}{jk}\right>,
\ee
where the indicator variables are defined the same way as above.  Splitting into two terms as before yields
\be
\left<B_iB_{jk}\right>=\sum_{\alpha}\left<\di{\alpha}{i}\di{\alpha}{jk}\right>+\sum_{\alpha\neq\beta}\left<\di{\alpha}{i}\right>\left<\di{\beta}{jk}\right>.
\ee
Once again, the first term will vanish unless $i=j$, leaving
\be
\left<B_iB_{jk}\right>=B_{jk}\delta^K_{ij}+\left(1-\frac{1}{N_{\rm vox}}\right)B_iB_{jk}.
\ee
and
\be
C_{ijk}^{\rm 1D\times2D}=B_{jk}\delta_{ij}^K-\frac{1}{N_{\rm vox}}B_iB_{jk}.
\ee

\subsection{Correlation between 1-D histograms}
The covariance between the two one-dimensional histogram of our two splits is somewhat more subtle.  Consider two extremes: If both maps are entirely noise dominated, the two histograms will be uncorrelated.  If both maps are entirely signal dominated the histograms will be identical (under our simplifying assumption that both observations map the same signal).  We need an expression that spans between both extremes.

Let us define two new indicators $s_\alpha(\delta)$ and $n_{\alpha}(\delta)$ such that $s$ or $n$ is unity if $\delta^S=\delta$ or $\delta^N=\delta$ for the signal and noise respectively.  We have defined these slightly differently since the binned $B_i$ are sourced by convolving continuous signal and noise values. We can write our previous indicator as
\be
\di{\alpha}{i}=\int_{\delta_i}\int s_\alpha(\delta')n_{\alpha}(\delta-\delta')d\delta'd\delta,
\ee
where the first integral is over the $i$'th histogram bin.  We also see that
$\langle s_{\alpha}(\delta)\rangle=\p^S(\delta)$, and similar for $n_{\alpha}$.

Since the signal and noise values are assumed to be independent, the expectation value of a histogram bin is
\be
\begin{aligned}
\left<B_i\right>&=\int_{\delta_i}\int\left<s_\alpha(\delta')\right>\left<n_{\alpha}(\delta-\delta')\right>d\delta'd\delta\\&=\int_{\delta_i}\int\p^S(\delta')\p^N(\delta-\delta')d\delta'd\delta.
\end{aligned}
\ee
This is just our original convolution between signal and noise.

Again writing out the first term of the covariance and separating terms where $\alpha=\beta$ we get
\be
\left<B_{1,i}B_{2,j}\right>=\sum_{\alpha}\left<\di{1,\alpha}{i}\di{2,\alpha}{j}\right>+\sum_{\alpha\neq\beta}\left<\di{1,\alpha}{i}\right>\left<\di{2,\beta}{j}\right>
\ee
The contribution to this sum from $\alpha=\beta$ is
\begin{multline}
\sum_{\alpha}\left<\di{1,\alpha}{i}\di{2,\alpha}{j}\right>=N_{\rm vox}\int_{\delta_i}\int_{\delta_j}\iint\left<s_\alpha(\delta')s_\alpha(\delta'')\right>\\ \times\p^N_1(\delta_a-\delta')\p^N_2(\delta_b-\delta'')d\delta'd\delta''d\delta_ad\delta_b.
\end{multline}
Since we have assumed identical signals, the remaining expectation value vanishes unless $\delta'=\delta''$.  This leaves
\begin{multline}
\sum_{\alpha}\left<\di{1,\alpha}{i}\di{2,\alpha}{j}\right>=N_{\rm vox}\int_{\delta_i}\int_{\delta_j}\int\p^S(\delta')\\ \times\p^N(\delta_a-\delta')\p^N(\delta_b-\delta')d\delta'd\delta_ad\delta_b,
\end{multline}
which is equal to the expected two-dimensional histogram $B_{ij}$.  Carrying out the rest of the algebra gives us our final covariance matrix
\be
C^{\rm 1D\times 1D}_{ij}=B_{ij}-\frac{1}{N_{\rm{vox}}}B_{1,i}B_{2,j}.
\ee

\section{The LIM Histogram}
\label{app:PofD}

Here we reproduce the derivation from \citet{Breysse2022bias} for the intensity mapping VID.  Assume each dark matter halo in the Universe with mass $M$ above some minimum mass $M_{\rm{min}}$ contains a point-source line emitter at its center.  Further assume that halos of a given mass have mean luminosity $L(M)$ with a lognormally-distributed scatter of width $\sigma_{\rm sc}$.  This model contains several simplifications of reality, neglecting among other things one-halo contributions from multiple sources within the same halo \citep{Schaan2021,Schaan2021a}, instrumental smearing and transfer functions \citep{esIV}, and intrinsic line widths \citep{Chung2021}.  However, it should capture the most important aspects of the signal.  We will express the observed intensity $T(\mathbf{x})$ as the sum of contributions from halos of a specific mass
\be
T(\mathbf{x})=\sum_iT(\mathbf{x}|M_i)dM,
\ee
where $T(\mathbf{x}|M_i)$ is the contribution to $T(\mathbf{x})$ from halos with masses between $M_i$ and $M_i+dM$.  As in Equations (\ref{pconv}--\ref{ft1d}), since we are summing the values for each mass bin we can write the overall intensity characteristic function $\ftp(\ft)$ as the product
\be
\ftp(\ft)=\prod_i\ftp(\ft|M_i),
\label{ftp_prod}
\ee
of the characteristic functions $\ftp(\ft|M_i)$ of halos with a specific mass.

Since mass is a continuous quantity, we are free to choose $dM$ to be as small as we desire.  Let us assume that $dM$ is small enough that each voxel contains exactly zero or one halo of a given mass. We can then write the intensity PDF in a single mass bin as
\be
\begin{aligned}
\p(T|M)&=\p(N_{\rm gal}=1)\p_1(T|M)+\p(N_{\rm gal}=0)\delta_D(T)\\
&=\p(N_{\rm gal}=1)\p_1(T|M)+\left(1-\p(N_{\rm gal}=1)\right)\delta_D(T),
\end{aligned}
\label{ptm}
\ee
where $\p_1(T|M)$ is the probability of observing intensity $T$ in a voxel known to contain exactly one object with mass between $M$ and $M+dM$.  Voxels which contain zero galaxies will obviously contribute zero intensity, leading to the Dirac delta function in the second term.  In the second equality, we used our assumption that each voxel contains no more than one such galaxy.  Fourier transforming Equation (\ref{ptm}) yields
\be
\ftp(\ft|M)=1+\p(N_{\rm gal}=1)\left(\ftp_1(\ft|M)-1\right),
\label{ptm1}
\ee
which is the characteristic function we need for Equation (\ref{ftp_prod}).  We stated above that we assume a lognormal scatter around some mean relation $L(M)$.  Under this assumption,
\be
\p_1(T|M)=\frac{1}{T\sigma_{\rm sc}\sqrt{2\pi}\ln(10)}\exp\left[-\frac{1}{2\sigma_{\rm sc}^2}\left(-\log(T)-\mu\right)^2\right],
\ee
where
\be
\mu\equiv\log\left(\frac{C_{LT}}{V_{\rm{vox}}}L(M)\right)-\frac{1}{2}\sigma_{\rm sc}^2\ln(10),
\ee
sets the mean of the distribution to the target value $L(M)$.

Assume that the galaxy number counts in our voxel are Poisson distributed around some expected mean $\overline{N}_{\rm gal}$.  Since our mass bin is arbitrarily small, we can say that $\overline{N}_{\rm gal}\ll1$.  This means that
\be
\p(N_{\rm gal}=1)=\overline{N}_{\rm gal}\exp\left[-\overline{N}_{\rm gal}\right]\approx \overline{N}_{\rm gal}.
\ee
We know that the halo abundance follows a mass function $dn/dM$, so we can write
\be
\p(N_{\rm gal}=1)=\frac{dn}{dM}V_{\rm vox}(1+b(M)\delta_m)dM.
\label{pNgal1}
\ee
In the absence of large-scale clustering, the term in parenthesis would be unity.  This term quantifies the possibility that our chosen voxel occupies either an over- or under-dense area of space.  We assume linear clustering where halos linearly trace the underlying dark matter field $\delta_m$ with bias $b(M)$.  The above expression is valid for a specific realization of the dark matter field, later we will average over all realizations to get the true distribution.

Combining Equations (\ref{ptm1}) and (\ref{pNgal1}) yields
\be
\begin{aligned}
\ftp(\ft|M)&=1+\frac{dn}{dM}V_{\rm vox}(1+b(M)\delta_m)\left(\ftp_1(\ft|M)-1\right)dM\\
&\equiv 1+p(\ft|M)(1+b(M)\delta_m)dM,
\end{aligned}
\ee
where we have compressed most of the mass dependence into $\p(\ft|M)$ for compactness.  Again knowing that $dM$ is small, we can use the Taylor expansion of the exponential to write
\be
\ftp(\ft|M)=\exp\left[p(\ft|M)(1+b(M)\delta_m)dM\right].
\ee
Plugging this into Equation (\ref{ftp_prod}) gives
\be
\begin{aligned}
\ftp(\ft)&=\prod_i\exp\left[p(\ft|M_i)(1+b(M_i)\delta_m)dM\right]\\
&=\exp\left[\sum_ip(\ft|M_i)(1+b(M_i)\delta_m)dM\right]\\
&=\exp\left[\int p(\ft|M)(1+b(M)\delta_m)dM\right],
\end{aligned}
\label{fullft}
\ee
where, since $dM$ is already infinitesimal, we have replaced the product of exponentials with the exponential of an integral.

We can separate Equation (\ref{fullft}) into two components.  The first,
\be
\ftp_{\rm un}(\ft)=\exp\left[\int p(\ft|M)dM\right],
\ee
gives the characteristic function in the absence of clustering, while the second
\be
\ftp_{\rm cl}(\ft)=\exp\left[\delta_m\int p(\ft|M)b(M)dM\right],
\ee
encodes the contribution from large-scale structure.  It is this second component that we need to average over all realizations of $\delta_m$.  Again assuming linear clustering, we use the identity that $\left<\exp(x)\right>=\exp(\left<x^2\right>/2)$ for Gaussian fields to get
\be
\ftp_{\rm{cl}}(\ft)=\exp\left[\frac{\sigma_m^2}{2}\left(\int p(\ft,M)b(M)dM\right)^2\right],
\ee
where
\be
\sigma_m^2=\langle\delta_m^2\rangle=\int P_m(\mathbf{k})\widetilde{W}_{\rm{vox}}^2(\mathbf{k})\frac{d^3\mathbf{k}}{(2\pi)^3},
\ee
is the variance of $\delta_m$ averaged over the voxel window function $W_{\rm{vox}})$.

We now have our final characteristic function
\be
\ftp(\ft)=\ftp_{\rm un}(\ft)\ftp_{\rm cl}(\ft).
\ee
We can inverse Fourier transform this quantity to get $\p(T)$ and $B_i$, or we can leave it as is for the DDE calculation.

\section{DDE dependence on model parameters}
\label{app:variation}

\begin{figure}
\centering
\includegraphics[width=\columnwidth]{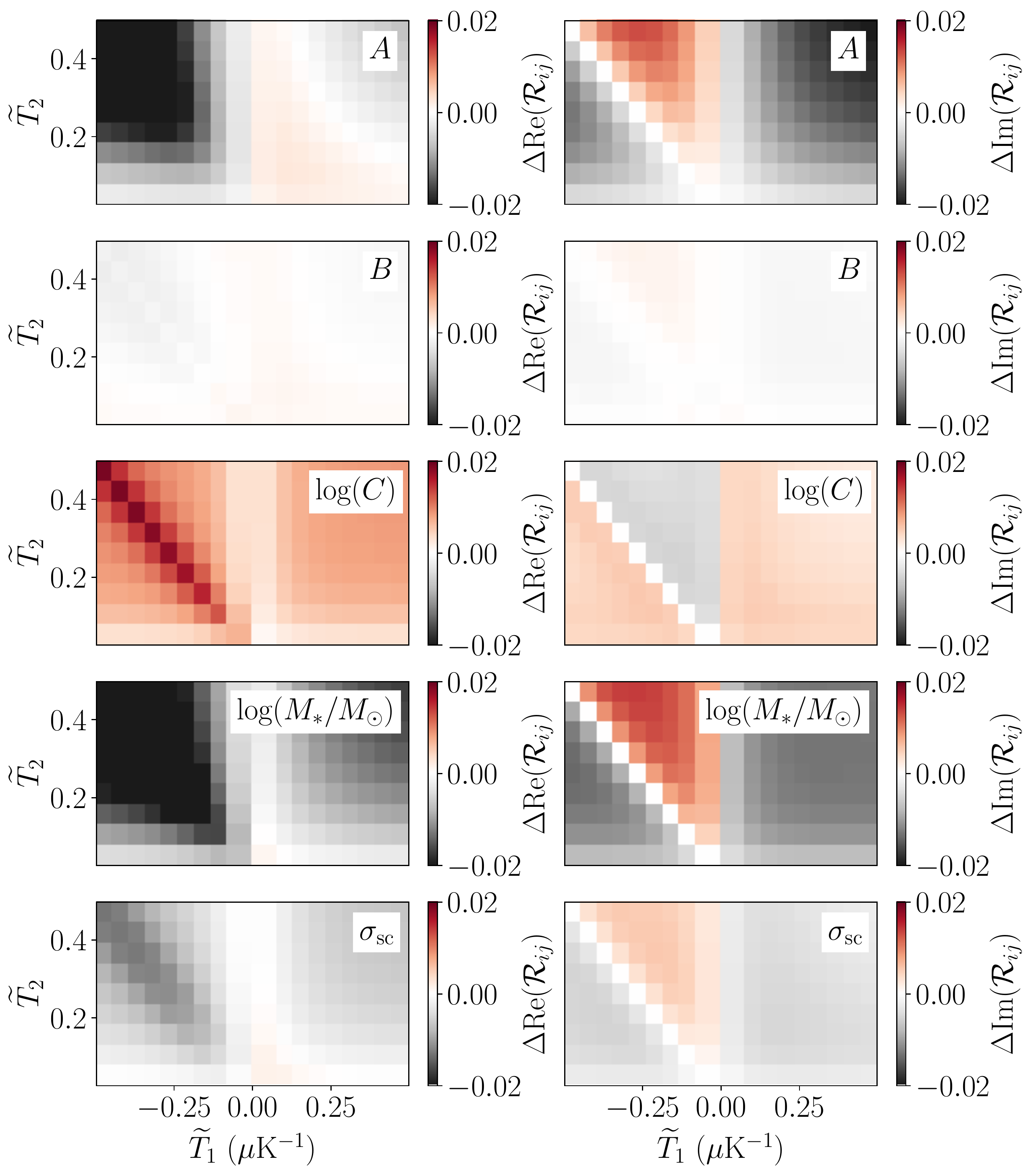}
\caption{Absolute change in the real part (left) and imaginary part (left) of the LIM DDE when increasing the labeled parameter by 25\%.}
\label{fig:params}
\end{figure}

\begin{figure}
\centering
\includegraphics[width=\columnwidth]{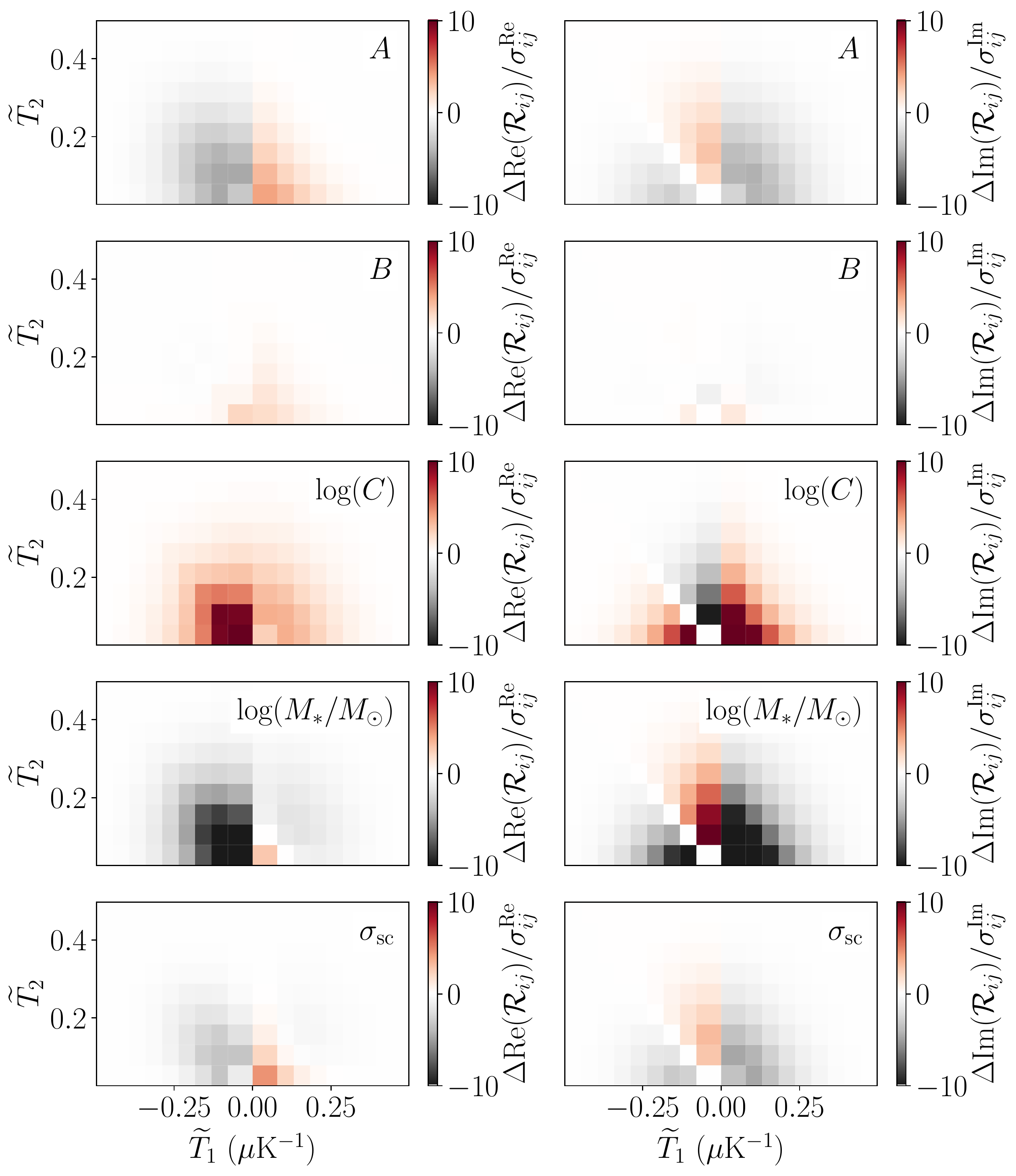}
\caption{Change in the real part (left) and imaginary part (left) of the LIM DDE when increasing the labeled parameter by 25\%, normalized by the 1-$\sigma$ error on $\rij$.}
\label{fig:variation}
\end{figure}

Figure \ref{fig:params} shows the absolute change in $\rij$ as the parameters of the $L(M)$ model from Equation (\ref{LofM}) are varied. Each panel shows the effect of increasing the relevant parameter by 25\%.  Figure \ref{fig:variation} shows the same, but normalized by the error on $\rij$.

\bsp	
\label{lastpage}
\end{document}